\documentclass[review,3p,times]{elsarticle}
\usepackage{amsmath,hyperref}
\usepackage{lineno}

\usepackage{epstopdf}
\journal{Communications in Nonlinear Science and Numerical Simulation}

\usepackage{lineno,hyperref}
\usepackage{graphicx}
\usepackage{dcolumn}
\usepackage{bm}
\usepackage{epsfig}
\usepackage{booktabs}
\usepackage{subfigure}
\usepackage{graphics}
\usepackage{graphicx}
\usepackage{amssymb}
\usepackage{amsmath}
\usepackage{array}
\usepackage{color}
\usepackage{booktabs}
\usepackage{multirow}
\usepackage{caption}
\usepackage{chngpage}
\usepackage{subfigure}
\usepackage{mathrsfs,gensymb,float}

\biboptions{numbers,sort&compress}
\modulolinenumbers[1]

\begin{document}

\captionsetup[figure]{labelfont={bf},name={Fig.},labelsep=period}        

\begin{frontmatter}
	
\title{ Modelling anisotropic Cahn-Hilliard equation with the lattice Boltzmann method}
\author[1]{Xinyue Liu}
\author[1]{Lei Wang\corref{mycorrespondingauthor}}
\cortext[mycorrespondingauthor]{Corresponding author}
\ead{wangleir1989@126.com}
\author[1]{Chenrui Liu}

\address[1]{School of Mathematics and Physics, China University of Geosciences, Wuhan 430074, China}


	
\begin{abstract}
The anisotropic Cahn-Hilliard equation is often used to model the formation of faceted pyramids on nanoscale crystal surfaces. In comparison to the isotropic Cahn-Hilliard model, the nonlinear terms associated with strong anisotropic coefficients present challenges for developing an effective numerical scheme. In this work, we propose a multiple-relaxation-time lattice Boltzmann method  to solve the anisotropic Cahn-Hilliard  equation. To this end, we reformulate the original  equation into a nonlinear convection-diffusion equation with source terms. Then the modified equilibrium distribution function and source terms are  incorporated into the computations. Through Chapman-Enskog  analysis, it successfully recovers the macroscopic governing equation. To validate the proposed approach, we perform numerical simulations, including cases like  droplet deformation and spinodal decomposition.   These results consistent with available works, confirming the effectiveness of the proposed approach. Furthermore, the simulations demonstrate that the model adheres to the energy dissipation law, further highlighting the effectiveness of the developed lattice Boltzmann method.

\end{abstract}

	
\end{frontmatter} 

\section{Introduction}
Multiphase flows play a vital role in both natural systems and engineering applications, such as oil and gas production, chemical processing, carbon dioxide enhanced oil recovery, and compressed air atomization \cite{Mouketou:cppm2019,Tokeshi:ac2002,Jiang:pecs2010}. A deep understanding the transport mechanisms in these systems is crucial for addressing the challenges associated with these processes.  In this setting, extensive experimental and theoretical researches have provided significant insights into multiphase flow transport over the years. \cite{Dudukovic:etfs2002,Yating:ciesc2024}.

Despite advances in experimental techniques and sophisticated instrumentation, there remain limitations and challenges in studying scenarios that are either experimentally inaccessible or prohibitively expensive and time-consuming. While theoretical analysis is valuable, it often fails to fully capture the complexity of real-world systems. With the rapid development of high-performance computing, numerical methods have become increasingly viable for exploring the intricate processes in multiphase flows. However, due to the complex interfacial dynamics between different phases, it is essential to accurately identify and track the interfaces between different fluids. Interface tracking or capturing techniques are a widely used numerical technique in multiphase flow modeling, with common methods include the  level set method \cite{Sussman:jcp1994}, the volume of fluid method \cite{Hirt:jcp1981,Raeini:jcp2012}, and the phase-field method \cite{Shen:siam2010}. However, the former two numerical methods often struggle with numerical stability, especially when calculating interfacial tension, and face challenges in simulating the internal structure of multiphase flows. The phase-field method, a diffuse interface approach, has gained attention due to its ability to describe interfaces more effectively. This method replaces sharp interfaces with thin and finite transition regions where properties like density and viscosity vary smoothly between fluids. It provides advantages in mass conservation and enhances the accuracy of interfacial curvature resolution.

The commonly used governing equation for the phase-field method is the Cahn-Hilliard (CH) equation \cite{Cahn:jcp1958}, and numerous researchers have developed efficient and accurate numerical models for the classical CH equation.  In 1998, Lowengrub et al. \cite{Lowengrub:prs1998} explored the dynamics of quasi-incompressible fluids governed by the CH equation, extending traditional models and thoroughly examining their broad applicability in simulating real-world multiphase flows. Badalassi et al. \cite{Badalassi:jcp2003} developed a phase-field model for density-matched binary fluids with variable mobility and viscosity, discretizing the CH equation semi-implicitly to overcome stability constraints caused by nonlinearity. Kim et al. \cite{Kim:jcp2004} considered the flow of two immiscible, incompressible, density-matched fluids, employing a nonlinear multigrid method to solve the CH equation. This approach can also be directly extended to multicomponent systems.  Gomez et al. \cite{Gomez:cmame2008} proposed an adaptive time-stepping method, presenting steady-state solutions in both two and three dimensions, thus paving the way for practical engineering applications of the phase-field method. For a more comprehensive overview, readers may refer to the review by Martin \cite{Worner:mn2012}  and the references therein. These studies primarily focused on isotropic CH equations. However, in nanoscale crystals, where the energy function  may become excessively large or singular in certain directions, the equilibrium shape may exhibit facets and corners, resulting in an anisotropic, multi-faceted pyramidal shape. To simulate such systems, recent studies have employed the phase-field method to solve anisotropic CH equation, incorporating large anisotropy coefficients to  capture the system’s anisotropy. When the anisotropy coefficient is reduced to zero, the system degenerates into the classical isotropic CH equation.Typically, analytical solutions for anisotropic CH equations are  unavailable, several effective numerical models have been proposed. Savina et al. \cite{Savina:pre2003} studied the faceting of crystal surfaces caused by strongly anisotropic surface tension, though their approach was restricted to long-wave approximations based on small variations in surface orientation.  Wise et al. \cite{Wise:jcp2007} introduced an efficient, second-order accurate, adaptive finite difference method for solving regularized, strongly anisotropic CH equations in two and three dimensions. Their approach discretized the anisotropic term in a conservative form in space and and used a fully implicit time integration. Torabi et al. \cite{Torabi:mpes2009} proposed an improved version of the phase-field model for strong anisotropy,  incorporating Willmore regularization. In this model, the square of the mean curvature is added to the energy functional to eliminate ill-posedness.  Chen et al. \cite{Chen:ccp2013} developed efficient and robust numerical methods for anisotropic Cahn-Hilliard systems, allowing for large time steps with stable time discretization and spectral discretization in space. Later, Chen et al. \cite{Chen:cmame2019} further refined these methods, employing a new stabilized Scalar Auxiliary Variable(SAV)  approach to eliminate oscillations caused by anisotropy coefficients, thereby enhancing stability while maintaining accuracy.

Following a different approach, the lattice Boltzmann (LB) method  has been widely applied in fluid dynamics, including multiphase flows. LB method effectively simulates multiphase flows by directly incorporating intermolecular interactions, offering advantages in boundary condition application and computational efficiency \cite{Fakhari:cnsns2009,Huang:2015,Li:pecs2016,Wang:cnsns2024}. Significant progress has also been made in phase-field-based LB method. He et al. \cite{He:jcp1999} first proposed an LB method for incompressible gas-liquid two-phase flows, capturing the phase interface using the order parameter $\phi$. Mukherjee and Abraham \cite{Mukherjee:pre2007} further advanced this by developing a two-phase LB method for fluid flow on a Cartesian grid, specifically addressing high gas-liquid density ratios. Lee et al. \cite{Lee:jcp2010} proposed an LB method for incompressible binary fluids, modeling contact line dynamics on partially wetting surfaces. Liang et al. \cite{Liang:pre2014} enhanced numerical stability with a multiple-relaxation-time LB method for incompressible multiphase flows, and Liang et al. \cite{Liang:cnsns2020} proposed an LB method for solving the fractional CH equation, integrating modified equilibrium distribution functions and appropriate source terms to simplify and effectively model complex systems with fractional derivatives.  Although these works have demonstrated the potential of phase-field-based LB method in multiphase flow simulations, they have primarily focused on classical models. However, as mentioned earlier, the CH equation can be anisotropic in some cases, which is relevant in many fields. To the best of our knowledge, no work in the LB community has addressed the simulation or solution of the anisotropic CH equation. To fill this gap, we propose the first LB method for the anisotropic CH equation. As noted in \cite{Torabi:mpes2009,Chen:ccp2013}, the anisotropic CH equation is essentially a nonlinear convection-diffusion equation. Compared to the isotropic CH equation, the anisotropic version contains terms with anisotropy parameters that can cause significant numerical oscillations. To address these challenges, we design an LB method that incorporates a modified equilibrium distribution function and source term into the current LB method to account for anisotropy. Additionally, we use a multiple-relaxation-time model to improve numerical stability. This model can address basic anisotropic two-phase problems and simulate complex two-phase transport processes.

The structure of this work is as follows. Section 2 provides the governing equation for anisotropic Cahn–Hilliard model. In Section 3, we present a new LB method  and Chapman-Enskog theory analysis. Section 4 tests the model on various examples, and Section 5 provides a brief conclusion.

\section{Macroscopic equations}

In this section, we derive the expression for the anisotropic Cahn-Hilliard  model using the phase-field method, which is a specific type of diffuse-interface approach based on the concept of free-energy functional. This method employs an order parameter $\phi$ to distinguish between two different phases: $\phi=1$ denotes one phase, and 
$\phi=-1$ denotes the other incompatible phase. At the interface between the two phases, $\phi$ takes the value $0$. Relying on previous research \cite{Torabi:mpes2009,Jacqmin:jcp1999}, the total free energy can be expressed as a function of the order parameter $\phi$:
\begin{equation}
	E(\phi)=\int_{\Omega}\mathcal{F}(\phi,\nabla\phi)d \Omega, \quad
	\mathcal{F}(\phi,\nabla\phi)=\gamma(\boldsymbol{n})(\frac{k}{2}|\nabla\phi|^2+F(\phi)),
	\label{eq1} 
\end{equation}
where $\mathcal{F}(\phi,\nabla\phi)$ is the free energy density  and $\Omega$ is  a regular and bounded region. $\gamma(\boldsymbol{n})$ is a function characterizing anisotropy, which will be 
expressed later. $\boldsymbol{n}$ is the interface normal and defined as $\mathbf{n}=\frac{\nabla\phi}{|\nabla\phi|}.$ Specifically, for two-dimensional (2D), it is  denoted as
\begin{equation}
	\mathbf{n}=(n_1,n_2)^T=\frac{1}{\sqrt{\phi_x^2+\phi_y^2}}(\phi_x,\phi_y)^T,
	\label{eq2}
\end{equation}
and for three-dimensional (3D), it is  written as
\begin{equation}
	\mathbf{n}=\left(n_{1},n_{2},n_{3}\right)^{T}=\frac{1}{\sqrt{\phi_{x}^{2}+\phi_{y}^{2}+\phi_{z}^{2}}}\left(\phi_{x},\phi_{y},\phi_{z}\right)^{T}.
	\label{eq3}
\end{equation}
Additionally, $k|\nabla\phi|^2/2$ represents the surface energy.
$F(\phi)$ denotes the bulk energy density, and takes the double-well potential form \cite{Lee:cpc2012}:
\begin{equation} 
	F(\phi)=\beta(\phi-1)^{2}(\phi+1)^{2},
	\label{eq4}
\end{equation}
where  $k$, $\beta$ are positive coefficients related to interfacial thickness $D$ and surface tension $\sigma$. They are given by \cite{Wang:C2019} 
\begin{equation}
 \sigma=\frac{4\sqrt{2k\beta}}{3},D=\sqrt{\frac{2k}{\beta}}.
\label{eq5}
\end{equation}

As for the anisotropic function $ \gamma(\boldsymbol{n})$  it can be expressed using the fourfold form \cite{Sekerka:JCG2005}  
\begin{equation}
\gamma(\boldsymbol{n})=1+\alpha\cos(4\theta)=1+\alpha\Big(4\sum_{i=1}^{d}n_{i}^{4}-3\Big),
	\label{eq6}
\end{equation}
where $\theta$ indicates the interface direction angle perpendicular to the interface and $\alpha$ is a non-negative constant indicating the strength of  anisotropy. In particular, when $\alpha = 0  \ (\text{i.e.,} \gamma(n) = 1)$, the system will degenerate into an isotropic state with the free energy being identical in all directions. However, the anisotropy of the system becomes more obvious for a relatively larger $\alpha$.

It is noted that the formation of Eq. (\ref{eq1}) is widely adopted in previous works \cite{Lee:cpc2012,Shen:MMAMS2012}. However, as pointed
out by these papers \cite{Torabi:mpes2009,Chen:ccp2013}, the system will become ill-posed if $\alpha$ is large enough.  To remedy this issue, the commonly used method is to introude a regularization term $G(\phi)$ into the original free energy, which can be expressed as  
\begin{equation}
	E(\phi)=\int_{\Omega}\left(\gamma(\boldsymbol{n})\left(\frac{k}{2}|\nabla\phi|^2+F(\phi)\right)+\frac{\epsilon}{2}G(\phi)\right)d\Omega,
	\label{eq7}
\end{equation}
where $\epsilon$ is a small regularization parameter.   In general,  the following two forms of regularization are considered \cite{Chen:cmame2019}.  One form of linear regularization is based on the bi-Laplacian of phase variable, which is expressed as
\begin{equation}
G(\phi)=(\Delta\phi)^2,
	\label{eq8}
\end{equation}
the other is the nonlinear Willmore regularization given by:
\begin{equation}
	G(\phi)=(\Delta\phi-f(\phi))^2,
		\label{eq9}
\end{equation}
where $f(\phi)=F'(\phi)=4\beta\phi(\phi+1)(\phi-1)$.

In the phase-field method, the evolution of  the order parameter $\phi$ is governed by  the well-known Cahn-Hilliard  equation, which is expressed as:
\begin{equation}
	\frac{\partial \phi}{\partial t}=\nabla\cdot(M\nabla\mu).
\label{eq10}
\end{equation}
Here, $M$ is  the mobility. $\mu$ is the chemical potential, which is  related to the interface driving energy. By introducing a variational operator  into the free energy functional Eq.(\ref{eq7}), the chemical potential $\mu$ can be obtained as:
\begin{equation}
	\mu=\frac{\delta E(\phi)}{\delta\phi}=\gamma(\boldsymbol{n})f(\phi)-\nabla\cdot\mathbf{Q}+\frac{\epsilon}{2}\frac{\delta G(\phi)}{\delta\phi}.	
	\label{eq11}
\end{equation}
Next, we present the specific forms of the above Eq.(\ref{eq11}) under different regularization conditions. They are given using the linear regularization:  
\begin{equation}
\mu=\gamma(\boldsymbol{n})f(\phi)-\nabla\cdot\mathbf{Q}+\epsilon\Delta^{2}\phi.	
	\label{eq12}
\end{equation}
or the Willmore regularization:
\begin{equation}
\mu=\gamma(\boldsymbol{n})f(\phi)-\nabla\cdot\mathbf{Q}+\epsilon(\Delta-f^{\prime}(\phi))(\Delta\phi-f(\phi)),
	\label{eq13}
\end{equation}
where $f'(\phi)=4\beta(3\phi^{2}-1)$.
The vector field $\mathbf{Q}$ is defined as
\begin{equation}
\mathbf{Q}=\gamma(\mathbf{n})k\nabla\phi+\frac{\mathbb{P}\nabla_{n}\gamma(\mathbf{n})}{|\nabla\phi|}(\frac{k}{2}|\nabla\phi|^{2}+F(\phi)),
	\label{eq14}
\end{equation}
where the matrix $\mathbb{P}=\mathbb{I}-\mathbf{n}\mathbf{n}^T.$ Since the presence of the expression $F(\phi)/|\nabla\phi|$  may lead to some problems in numerical calculations,  the following approximation near the interface  is used to simplify the calculation \cite{Chen:ccp2013,Torabi:mpes2009},
	$F(\phi)\sim{k}|\nabla\phi|^{2}/2$.
In such a case, the vector field  can be rewritten as 
\begin{equation}
\mathbf{Q}\approx\gamma(\boldsymbol{n})k\nabla\phi+|\nabla\phi|\mathbb{P}\nabla_{n}\gamma(\boldsymbol{n}). 
	\label{eq15}
\end{equation}

\section{Lattice Boltzmann method}
In this section, we intend to propose an LB method for the anisotropic Cahn–Hilliard equation. Unlike the isotropic Cahn-Hilliard equation \cite{Wang:C2019}, the anisotropic Cahn-Hilliard equation contains anisotropic terms, which pose challenges in developing the corresponding LB method. 
To address this issue, we reformulate the anisotropic equation (\ref{eq10}) into the following form:
\begin{equation}
	\begin{aligned}
		&\frac{\partial \phi}{\partial t}=\nabla\cdot(M\nabla\mu_\alpha)+\nabla\cdot(M\nabla\mu_\beta),\\
		&\mu_\alpha=f(\phi)-k\nabla^{2}\phi,\\
		&\mu_\beta=\mu-\mu_\alpha.\\
	\end{aligned}
	\label{eq16}
\end{equation}
Comparing the anisotropic and isotropic Cahn-Hilliard equations, it is evident that when $\mu_\beta=0$ (i.e., $\alpha=0$), the equation reduces to the isotropic case. Here, Equation (\ref{eq16}) is considered a convection-diffusion equation with the source term $\nabla\cdot(M\nabla\mu_\beta)$. Based on the collision operator, the LB  method  is categorized into the  single-relaxation-time (SRT) model \cite{He:pre1997}, the two-relaxation-time (TRT) model \cite{Ginzburg:ccp2008} and the
multiple-relaxation-time (MRT) model \cite{Lallemand:pre2000}. Due to the complexity of the anisotropic terms, which may impact numerical stability, the MRT model is employed in the current work, and extensive previous have proven that the MRT model shares a better numerical stability over the SRT model \cite{Luo:pre2011}. 

The LB equation with MRT collision operator can be expressed as  
\begin{equation}
	{f}_{i} \left ( \mathbf{x} +\mathbf{c}_{i}\delta _{t},t+\delta _{t}\right )- {f}_{i}  \left ( \mathbf{x} ,t\right )=-(\mathbf{M}^{-1}\mathbf{S}\mathbf{M})_{ij}\left [  f_{j} \left ( \mathbf{x} ,t\right)-  {f}_{j}^{(\mathrm{eq}) }  \left (\mathbf{x} ,t\right)  \right ]+ \delta _{t} [\mathbf{M}^{-1}({\mathbf{I}}-\frac{{\mathbf{S}}}{2})\mathbf{M}]_{ij}{F}_{j}. 
		\label{eq17} 
\end{equation}
where $\delta_{t}$ is the time step and $f_{i}\left(\mathbf{x} ,t\right)$ ($i=0,1,2,\cdots,q-1$ with $q$ representing the number of discrete velocity directions)  denotes the distribution function of order parameter $\phi$. $f_{i}^{(eq)}$ is the  equilibrium distribution function at position $\mathbf{x}$ and time $t$, and according to the work \cite{Chai:siam2019,Bao:pre2024}, it can be written as
\begin{equation}
f_i^{(eq)}=\begin{cases}\phi+(\omega_i-1)\mu_\alpha,\quad&i=0,\\\omega_i\mu_\alpha,\quad&i\neq0.\end{cases}
\label{eq18} 
\end{equation}
$F_i$ is the distribution function of the source term, and is defined by
\begin{equation}
	F_i=-\omega_i \mathbf{c}_i\cdot \nabla\mu_\beta.
	\label{eq19} 
\end{equation}
Here, the weight coefficients $\omega_i$  and the discrete velocities $\mathbf{c}_i$ are given for the $DdQq$ discrete-velocity model ($q$ discrete velocities in d-dimensional space, $d = 1, 2, 3$) as follows:

$D2Q9$: 
\begin{subequations}
	\begin{align}	
		\mathbf{c}_i=&\begin{bmatrix}0&1&0&-1&0&1&-1&-1&1\\0&0&1&0&-1&1&1&-1&-1\end{bmatrix}c,\tag{{20}{a}}\label{20a}\\
		&\omega_{0}=4/9,\omega_{1-4}=1/9,\omega_{5-8}=1/36;\tag{{20}{b}}\label{20b}
	\end{align}
	\label{eq20}
\end{subequations}

$D3Q15$: 
\setcounter{MaxMatrixCols}{20}
\begin{subequations}
	\begin{align}	
		\mathbf{c}_i=&\left[\begin{array}{rrrrrrrrrrrrrrr}0&1&-1&0&0&0&0&1&-1&1&-1&1&-1&-1&1\\0&0&0&1&-1&0&0&1&-1&1&-1&-1&1&1&-1\\0&0&0&0&0&1&-1&1&-1&-1&1&1&-1&1&-1\end{array}\right]c,\tag{{21}{a}}\label{21a}\\
		&\omega_{0}=2/9,\omega_{1-6}=1/9,\omega_{7-14}=1/72,\tag{{21}{b}}\label{21b}
	\end{align}
	\label{eq21}
\end{subequations}
where $c=\delta_{x}/\delta_{t}$ ($\delta_{x}$ is the lattice spacing) and $c_s=c /\sqrt{3}$ is the sound speed. In this setting, the order parameter $\phi$ is computed by
\begin{equation}
 \phi=\sum_i f_i
 \label{eq22}
\end{equation}
 The  $q \times q$ transformation matrix $\mathbf{M}$  in Eq. (\ref{eq17})  is constructed from the discrete velocities via the Gram-Schmidt orthogonalization procedure \cite{dHumieres:p2002}. For the D2Q9 discrete velocity model, $\bf{M}$ can be chosen as 
\begin{equation}
	\mathbf{M}=\left(\begin{array}{ccccccccc}
		1 & 1 & 1 & 1 & 1 & 1 & 1 & 1 & 1 \\
		-4 & -1 & -1 & -1 & -1 & 2 & 2 & 2 & 2 \\
		4 & -2 & -2 & -2 & -2 & 1 & 1 & 1 & 1 \\
		0 & 1 & 0 & -1 & 0 & 1 & -1 & -1 & 1 \\
		0 & -2 & 0 & 2 & 0 & 1 & -1 & -1 & 1 \\
		0 & 0 & 1 & 0 & -1 & 1 & 1 & -1 & -1 \\
		0 & 0 & -2 & 0 & 2 & 1 & 1 & -1 & -1 \\
		0 & 1 & -1 & 1 & -1 & 0 & 0 & 0 & 0 \\
		0 & 0 & 0 & 0 & 0 & 1 & -1 & 1 & -1
	\end{array}\right),
	\label{eq23}
\end{equation}
while for the D3Q19 lattice, the orthogonal transformation matrix $\bf{M}$ can be expressed as 
\setcounter{MaxMatrixCols}{20}
\begin{equation}
	\mathbf{M}=\begin{pmatrix}1&1&1&1&1&1&1&1&1&1&1&1&1&1&1\\
		-2&-1&-1&-1&-1&-1&-1&1&1&1&1&1&1&1&1\\
		16&-4&-4&-4&-4&-4&-4&1&1&1&1&1&1&1&1\\
		0&1&-1&0&0&0&0&1&-1&1&-1&1&-1&1&-1\\
		0&-4&4&0&0&0&0&1&-1&1&-1&1&-1&1&-1\\
		0&0&0&1&-1&0&0&1&1&-1&-1&1&1&-1&-1\\
		0&0&0&-4&4&0&0&1&1&-1&-1&1&1&-1&-1\\
		0&0&0&0&0&1&-1&1&1&1&1&-1&-1&-1&-1\\
		0&0&0&0&0&-4&4&1&1&1&1&-1&-1&-1&-1\\
		0&2&2&-1&-1&-1&-1&0&0&0&0&0&0&0&0\\
		0&0&0&1&1&-1&-1&0&0&0&0&0&0&0&0\\
		0&0&0&0&0&0&0&1&-1&-1&1&1&-1&-1&1\\
		0&0&0&0&0&0&0&1&1&-1&-1&-1&-1&1&1\\
		0&0&0&0&0&0&0&1&-1&1&-1&-1&1&-1&1\\
		0&0&0&0&0&0&0&1&-1&-1&1&1&-1&1&-1\end{pmatrix}.
		\label{eq24}
\end{equation}
By using $M$, the distribution function $f_{i}$, the equilibrium distribution function $f_{i}^{(e q)}$ and the soure term $F_{i}$ can be projected from the discrete  space onto macroscopic variables in the moment space \cite{Lallemand:pre2000} through following relations
\begin{equation}
	\mathbf{m}=\mathbf{M}\mathbf{f} \left ( \mathbf{x} ,t\right),{\mathbf{m}}^{(\mathrm{eq})}=\mathbf{M}{\mathbf{f}}^{(\mathrm{eq})} \left ( \mathbf{x} ,t\right),{\mathbf{F}}_{m}=\mathbf{M}\mathbf{F},
		\label{eq25}
 \end{equation}
in which ${\mathbf{f}}\left ( \mathbf{x} ,t\right)=\left[f_{0}\left(\mathbf{x} ,t\right), \cdots, f_{q-1}\left(\mathbf{x} ,t\right)\right]^{\mathrm{T}}$,  ${\mathbf{f}}^{(\mathrm{eq})} \left ( \mathbf{x} ,t\right)=\left[f_{0}^{(\mathrm{eq})}\left(\mathbf{x} ,t\right), \cdots, f_{q-1}^{(\mathrm{eq})}\left(\mathbf{x} ,t\right)\right]^{\mathrm{T}}$ and $\mathbf{F}=\left[F_{0}, \cdots, F_{q-1} \right]^{\mathrm{T}}$. The relaxation matrix in the moment space is given by $\mathbf{S}=\mathrm{diag}\left(s_{0},s_{1},s_{2}, \cdots ,s_{q-1}\right)$, in which the diagonal element $s_{i}$ is the relaxation parameter corresponding to the $i$th moment $m_i$.

In practice, it is noted that there are some derivative terms in the LB evolution equations, which should be discretized with suitable difference schemes. To this end, the second-order isotropic difference schemes \cite{Guo:pre2011,Lou:el2012} to compute the gradient term and the Laplace operator,
\begin{subequations}
	\begin{align}
		\nabla\chi\left(\mathbf{x},t\right)&=\sum_{i}\frac{\omega_{i}\mathbf{c}_{i}\chi\left(\mathbf{x}+\mathbf{c}_{i}\delta_{t},t\right)}{c_{s}^{2}\delta_{t}}, \tag{{26}{a}}\label{26a} \\
		\nabla^{2}\chi\left(\mathbf{x},t\right)&=\sum_{i}\frac{2\omega_{i}[\chi\left(\mathbf{x}+\mathbf{c}_{i}\delta_{t},t\right)-\chi\left(\mathbf{x},t\right)]}{c_{s}^{2}\delta_{t}^{2}},\tag{{26}{b}}\label{26b}
	\end{align}
	\label{eq26}
\end{subequations}
where $\chi$ represents an arbitrary variable.

In what follows, we turn to present the Chapman-Enskog (CE) analysis to drive the macroscopic governing equation Eq. (\ref{eq16}) from the present LB method, using $D2Q9$ discrete-velocity model as an example. To this end, we first recall the definition of the weight coefficient $\omega_i$, equilibrium distribution function  $f_i^{(eq)}$
and forcing term $F_i$, we can easily obtain the following moments,
\begin{subequations}
	\begin{align}
			&\sum_{i}\omega_{i}=1,\quad\sum_{i}\mathbf{c}_{i}\omega_{i}=0,\quad\sum_{i}\mathbf{c}_{i}\mathbf{c}_{i}\omega_{i}=c_{s}^{2}\mathbf{I},\tag{{27}{a}}\label{27a}\\
		&\sum_i f_i^{(eq)} =\phi, \quad\sum_i \mathbf{c}_i f_i^{(eq)}=0, \quad\sum_i \mathbf{c}_i \mathbf{c}_if_i^{(eq)}= c_s^2\mu_\alpha\mathbf{I},\tag{{27}{b}}\label{27b} \\
		&\sum_iF_i=0,\quad\sum_i\mathbf{c}_iF_i=-c_s^2\nabla\mu_\beta.\tag{{27}{c}}\label{27c}
	\end{align}
	\label{eq27}
\end{subequations}
Based on the CE analysis,  the distribution function $f_i$, the source term $F_i$ and the derivatives of time and space can be expressed as 
\begin{equation}
	\begin{aligned}
		f_i &=\sum_{n=0}^{+\infty} \varepsilon^{n} {f_i}^{(n)}, \quad {\mathbf{F}}_{m}=\varepsilon {\mathbf{F}}_{m}^{(1)},\quad F_i=\varepsilon F_i^{(1)}.\\
		\partial_{t} &=\varepsilon \partial_{t 1}+\varepsilon^{2} \partial_{t 2}, \quad \nabla=\varepsilon \nabla_{1} 
	\end{aligned}
	\label{eq28}
\end{equation}
where  $\varepsilon$ is a small expansion parameter.

Firstly, applying the Taylor expansion to Eq. (\ref{eq17}), we can obtain
\begin{equation}
{D}_i f_i +\frac{\delta_{t}}{2} {{D}_i}^{2} f_i+O\left(\delta_{t}^{2}\right)=-\frac{1}{\delta t}(\mathbf{M}^{-1}\mathbf{S}\mathbf{M})_{ij} [f_j-  {f_j}^{(\mathrm{eq}) }]  +[\mathbf{M}^{-1}({\mathbf{I}}-\frac{{\mathbf{S}}}{2})\mathbf{M}]_{ij}{F}_{j}.  
	\label{eq29}
\end{equation}
Here, ${D}_i= \partial_{t}+\mathbf{c}_{i} \cdot \nabla$. Next, substituting Eq. (\ref{eq28}) into Eq. (\ref{eq29}), we can obtain the following equations at the different orders of $\varepsilon$: 
\begin{subequations}
	\begin{align}
		&\varepsilon^{0}:\ f_{i}^{(0)}=f_{i}^{(eq)}, \tag{{30}{a}}\label{30a}\\
		&\varepsilon^{1}:\ D_{1i}f_{i}^{(0)}=-\frac{1}{\delta t}(\mathbf{M}^{-1}\mathbf{SM})_{ij}f_{j}^{(1)}+[\mathbf{M}^{-1}({\mathbf{I}}-\frac{{\mathbf{S}}}{2})\mathbf{M}]_{ij}F_{j}^{(1)}, \tag{{30}{b}}\label{30b}\\
		&\varepsilon^{1}:\ \partial_{t_{2}}f_{i}^{(0)}+D_{1i}f_{i}^{(1)}+\frac{\delta t}{2}D_{1i}^{2}f_{i}^{(0)}=-\frac{1}{\delta t}(\mathbf{M}^{-1}\mathbf{SM})_{ij}f_{j}^{(2)},\tag{{30}{c}}\label{30c}
	\end{align}
\label{eq30}
\end{subequations}
where ${D}_{1i}= \partial_{t_1}+\mathbf{c}_{i} \cdot \nabla_1$.  Then, we substitute Eq. \eqref{30b} into Eq. \eqref{30c} and simplify Eq. \eqref{30c} with the following form,
\begin{equation}
	\partial_{t_2} f^{(0)}+-\frac{1}{\delta t}D_{1i}\left(\mathbf{I}-\frac{\mathbf{M}^{-1}\mathbf{SM}}{2}\right)_{ij}({f_j}^{(1)}+\frac{\delta_{t}}{2}{{F}}_{j}^{(1)})
	=-\frac{1}{\delta_{t}}(\mathbf{M}^{-1}\mathbf{SM})_{ij} {f_j}^{(2)}.
	\label{eq31}
\end{equation}
Rewriting Eqs. (\ref{eq31}) in a vector form and multiplying $\mathbf{M}$ on both sides of them, we can rewrite them in the moment space,
\begin{subequations}
	\begin{align}
		\varepsilon^{0}:&\ \mathbf{m}^{(0)}=\mathbf{m}^{(\mathrm{eq})}, \tag{{32}{a}}\label{32a}\\ 
		\varepsilon^{1}:&\ \left(\mathbf{I} \partial_{t_ 1}+\mathbf{E} \cdot \nabla_{1}\right) \mathbf{m}^{(0)}-{\mathbf{F}}_{m}^{(1)}=-\frac{\mathbf{S}}{\delta_{t}}(\mathbf{m}^{(1)}+\frac{\delta_{t}}{2}{\mathbf{F}}_{m}^{(1)}) \tag{{32}{b}}\label{32b}\\ 
		\varepsilon^{2}:& \ \partial_{t_ 2} \mathbf{m}^{(0)}+\left(\mathbf{I} \partial_{t_1}+\mathbf{E} \cdot \nabla_{1}\right)\left(\mathbf{I}-\frac{\mathbf{S}}{2}\right)(\mathbf{m}^{(1)}+\frac{\delta_{t}}{2}{\mathbf{F}}_{m}^{(1)})
		=-\frac{\mathbf{S}}{\delta_{t}} \mathbf{m}^{(2)}. \tag{{32}{c}}\label{32c}
	\end{align}
	\label{eq32}
\end{subequations}
in which $\mathbf{E}=\left(\mathbf{E}_{x}, \mathbf{E}_{y}\right)^{\mathrm{T}}$,  $\mathbf{E}_{x}=\mathbf{M}\left[\operatorname{diag}\left(c_{0 x}, \ldots, c_{8 x}\right)\right] \mathbf{M}^{-1}$ and $\mathbf{E}_{y}=\mathbf{M}\left[\operatorname{diag}\left(c_{0 y}, \ldots, c_{8 y}\right)\right] \mathbf{M}^{-1}$. 
Based on Eqs. (\ref{eq32}),  we
separately write the equations at the different orders of $\varepsilon$ and present the first one because these three equations are necessary for deriving the anisotropic CH equation,
\begin{subequations}
\begin{align}
	\varepsilon^{0}:&\ m_{0}^{(0)}=m_{0}^{(\mathrm{eq})},\tag{{33}{a}}\label{33a} \\
	\varepsilon^{1}: &\ \partial_{t 1} m_{0}^{(0)}+c\nabla_{1} \cdot\left(\begin{array}{c}
		m_{3}^{(0)} \\
		m_{5}^{(0)}
	\end{array}\right)-F_0^{(1)}=-\frac{s_{0}}{\delta_{t}}(m_{0}^{(1)}+\frac{\delta t}{2}F_0^{(1)}), \tag{{33}{b}}\label{33b}\\
	\varepsilon^{2}:&\ \partial_{t 2} m_{0}^{(0)}+\partial_{t 1}\left(1-\frac{s_{0}}{2}\right)m_{0}^{(1)} 
	+c\nabla_{1} \cdot\left(1-\frac{s_{3,5}}{2}\right)\left(\begin{array}{l}
		m_{3}^{(1)} \\
		m_{5}^{(1)}
	\end{array}\right)+c\nabla_{1} \cdot\left(1-\frac{s_{3,5}}{2}\right)\frac{\delta t}{2}\left(\begin{array}{l}
	F_{3}^{(1)} \\
	F_{5}^{(1)}
\end{array}\right)=-\frac{s_{0}}{\delta_{t}} m_{0}^{(2)}.\tag{{33}{c}}\label{33c}
\end{align}
\label{eq33}
\end{subequations}
In addition, by Eqs. (\ref{eq27}) and \eqref{33a}, the constraint conditions for the conserved moment $m_{0}$ can be expressed as 
\begin{equation}
	m_{0}^{(k)}=0(\forall k \geqslant 1).
	\label{eq34}
\end{equation}
Therefore, Eqs. \eqref{33b} and \eqref{33c} can be simplified as
\begin{subequations}
	\begin{align}
		\varepsilon^{1}:& \partial_{t 1} \phi=0,\tag{{35}{a}}\label{35a}\\
		\varepsilon^{2}:&\partial_{t 2} \phi+ c\nabla_{1} \cdot\left(1-\frac{s_{3,5}}{2}\right)\left(\begin{array}{c}
			m_{3}^{(1)} \\
			m_{5}^{(1)}
		\end{array}\right)+c\nabla_{1} \cdot\left(1-\frac{s_{3,5}}{2}\right)\frac{\delta t}{2}\left(\begin{array}{l}
		F_{3}^{(1)} \\
		F_{5}^{(1)}
	\end{array}\right)=0.\tag{{35}{b}}\label{35b}
	\end{align}
	\label{eq35}
\end{subequations}
With the further aid of the first-order Eq. \eqref{32b} in $\varepsilon$,  we can obtain the fourth and sisth ones to  simplify Eq. \eqref{35b},
\begin{equation}
	\begin{aligned}
		-\frac{s_{j}}{\delta_{t}}\left(\begin{array}{l}
			m_{3}^{(1)} \\ 
			m_{5}^{(1)}
		\end{array}\right) 
		=\begin{pmatrix}\partial_{t1}m_{3}^{(0)}+\frac{2c}{3}\partial_{x1}m_{0}^{(0)}+\frac{c}{6}\partial_{x1}m_{1}^{(0)}+\frac{c}{2}\partial_{x1}m_{7}^{(0)}+c\partial_{y1}m_{8}^{(0)}+\frac{s_3}{2}F_3^{(1)}- F_3^{(1)}\\ \partial_{t1}m_{5}^{(0)}+c\partial_{x1}m_{8}^{(0)}+\frac{2c}{3}\partial_{y1}m_{0}^{(0)}+\frac{c}{6}\partial_{y1}m_{1}^{(0)}-\frac{c}{2}\partial_{y1}m_{7}^{(0)}+\frac{s_5}{2}F_5^{(1)}-F_5^{(1)}\end{pmatrix}=\frac{c}{3} \nabla_{1} \mu_\alpha +\frac{c}{3}(\frac{s_{3,5}}{2}-1) \nabla_{1}\mu_\beta.
	\end{aligned}
	\label{eq36}
\end{equation}
Thus Eq. \eqref{35b} can be written as
\begin{equation}
	\varepsilon^{2}:\partial_{t 2}\phi= \nabla_{1} \cdot\left[c_s^2\left(\frac{1}{s_{j}}-\frac{1}{2}\right) \delta_{t} \nabla_{1} \mu_\alpha\right]+\nabla_{1} \cdot\left[c_s^2\left(\frac{1}{s_{j}}-\frac{1}{2}\right) \delta_{t} \nabla_{1} \mu_\beta\right].
	\label{eq37}
\end{equation}
Summing Eq. \eqref{35a} and Eq. (\ref{eq36}), the governing equation can be recovered,
\begin{equation}
	\partial_{t }\phi=\nabla \cdot\left[c_s^2\left(\frac{1}{s_{3,5}}-\frac{1}{2}\right) \delta_{t} \nabla (\mu_\alpha+\mu_\beta)\right].
	\label{eq38}
\end{equation}
Lastly, comparing Eq. (\ref{eq38}) with Eq. (\ref{eq16}), the diffusion coefficient is obtained as $M=c_s^2(1/s_{3,5}-0.5)\delta_{t}$ with $s_3=s_5$.

\section{Numerical results and discussions }
 In this section, we first perform a grid independence test for the lattice Boltzmann method applied to the 2D anisotropic Cahn-Hilliard equation. Then, numerical tests include the anisotropic envolution of one and two droplets  and spinodal decomposition in 2D are carried out to further verify the present model. After that, to show the feasibility of the proposed LB method for some high-dimensional cases, we investigate anisotropic envolution of one and two droplets in 3D. Referring to previous work \cite{Chen:ccp2013}, we find that the Willmore  regularization performs better numerically and all their edges and corners are closer to the asymptotic results \cite{Spencer:pre2004} than the linear regularization. Therefore in the current work we use Willmore regularization for numerical simulation. Unless otherwise stated,  the boundary conditions are periodic boundaries \cite{Maier:pof1996} and the   relevant parameters values are set as follows:
 
\begin{equation}
	M=0.1,\ \epsilon=0.03,\ D=5.0,\ \sigma=0.01.
\label{eq39}
\end{equation}

\begin{figure}[htbp]
	\centering
	\subfigure[]{\includegraphics[width=0.29\textwidth]{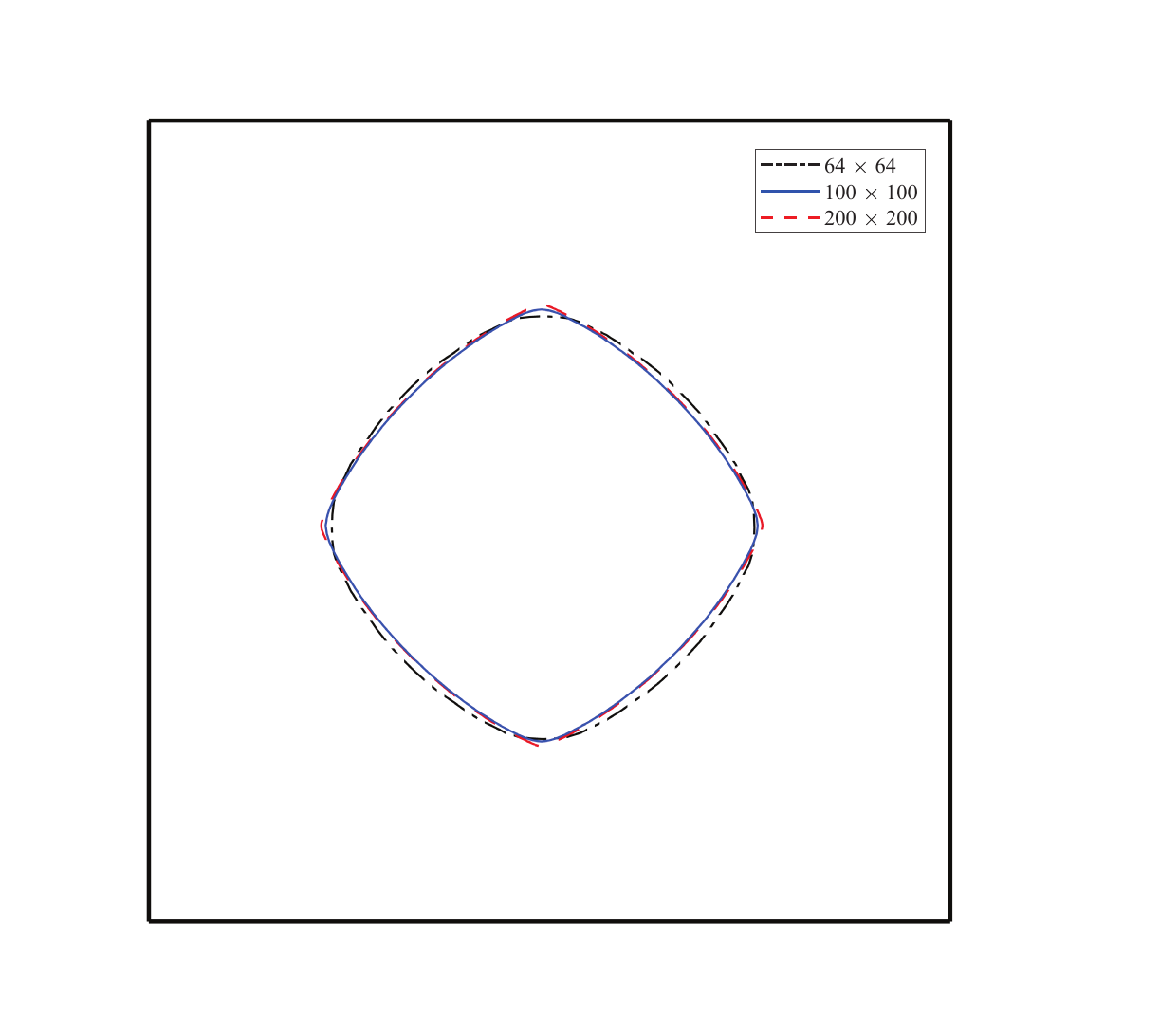}}\label{fig1(a)}\quad\quad\quad\quad
	\subfigure[]{\includegraphics[width=0.32\textwidth]{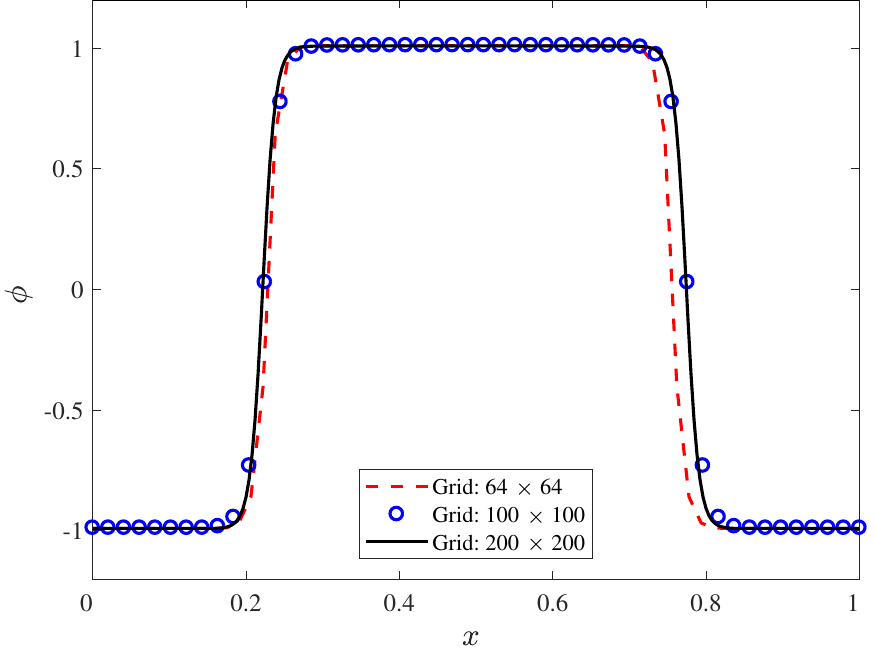}}\label{fig1(b)}
	\caption{The distribution of the phase variable $\phi$ along  the mid-height of the $y=H/2.0$ at different grid resolutions}
	\label{fig1}
\end{figure}

In addition, we perform a numerical study involving grid  independence tests  in the anisotropic Cahn-Hilliard system. More specifically,  we compare three different grid sizes: $64 \times 64$, $100 \times 100$, and $200 \times 200$. The numerical results are shown in Fig. \ref{fig1}.  As one can see,  the order parameter $\phi$ distribution obtained using the $100 \times 100$ grid closely matches that of the $200 \times 200$ grid, while slight discrepancies are noted with the $64 \times 64$ grid.  This is because a smaller grid  grid brings some numerical precision errors. In addition, we note that the computation time for the $100 \times 100$ grid is significantly shorter compared to the $200 \times 200$ grid. Therefore, considering both computational efficiency and numerical accuracy, we select the $100 \times 100$ grid for further computations.

\subsection{Anisotropic evolution   for single  droplet in 2D}
\begin{figure}[H]
	\centering
	\subfigure[$\alpha$=0.1]{\includegraphics[width=0.2\textwidth]{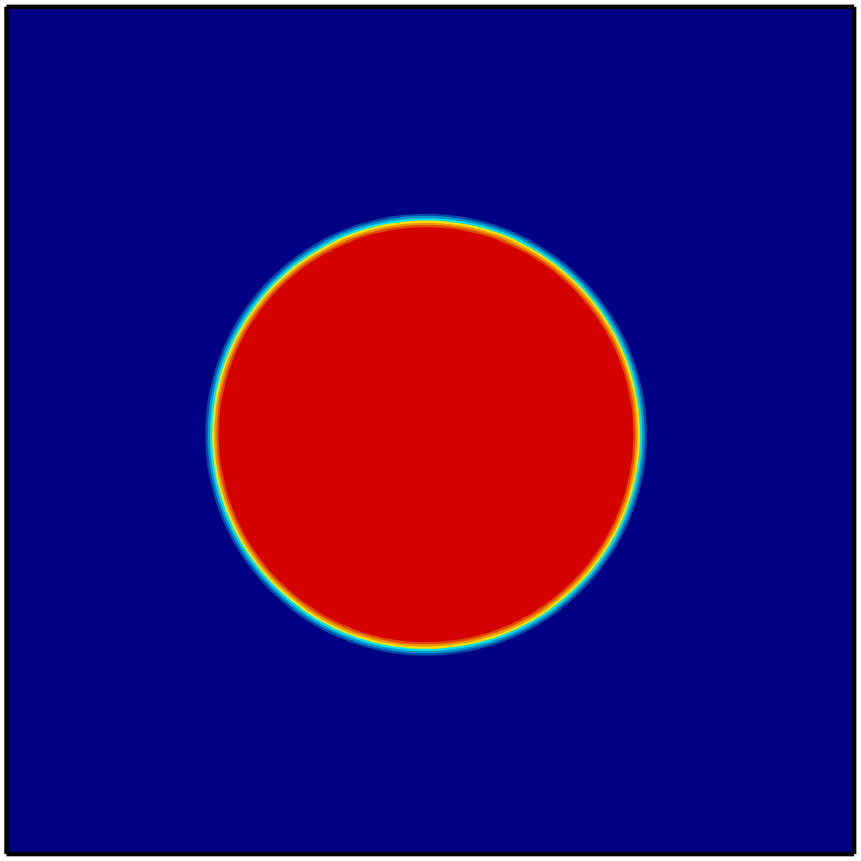}\quad\quad\includegraphics[width=0.2\textwidth]{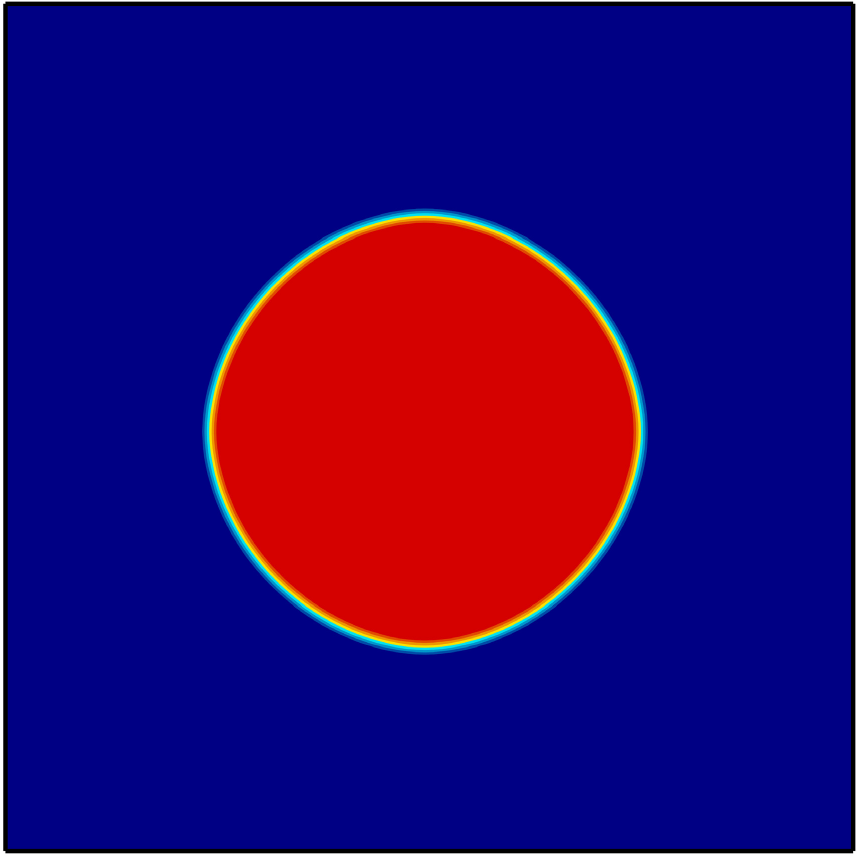}\quad\quad\includegraphics[width=0.2\textwidth]{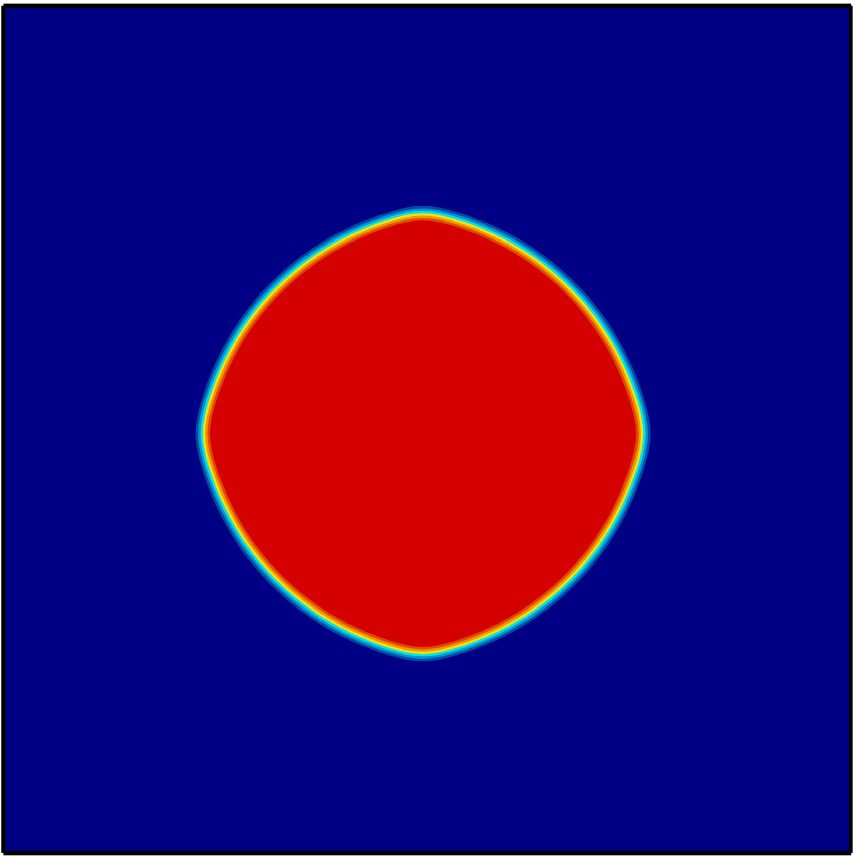}\quad\quad\includegraphics[width=0.2\textwidth]{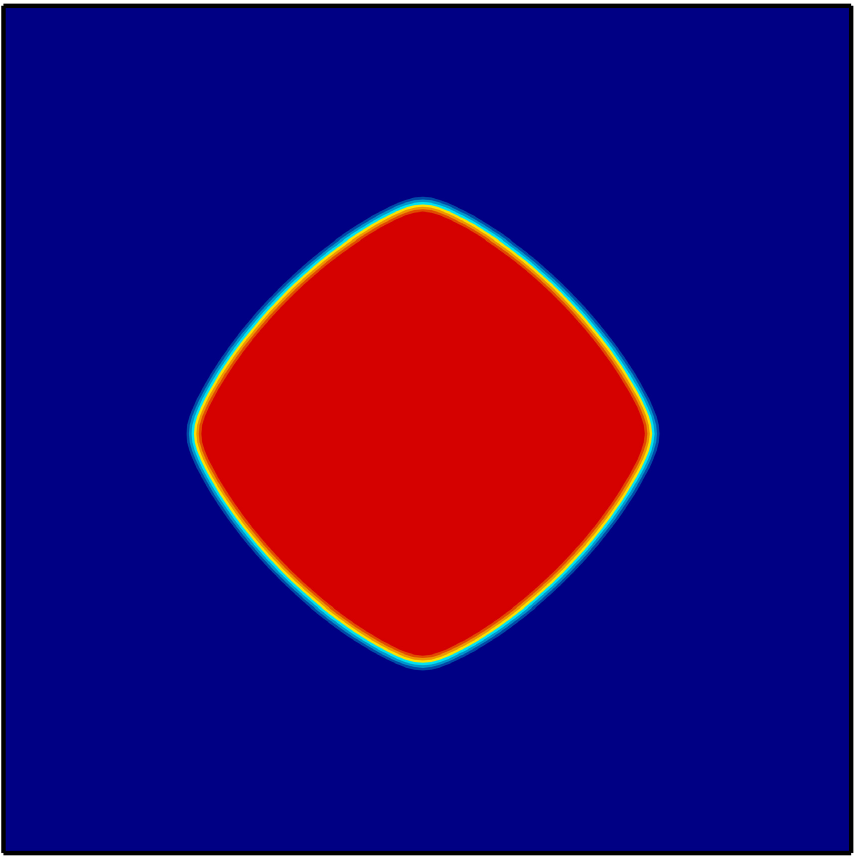}}
	\subfigure[$\alpha$=0.2]{\includegraphics[width=0.2\textwidth]{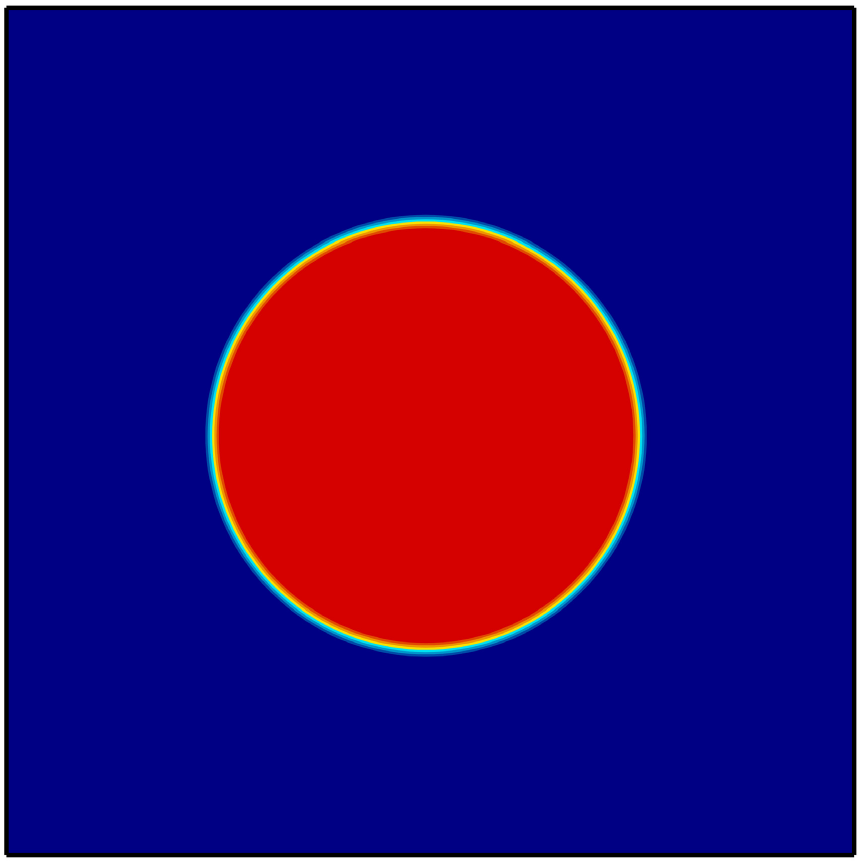}\quad\quad\includegraphics[width=0.2\textwidth]{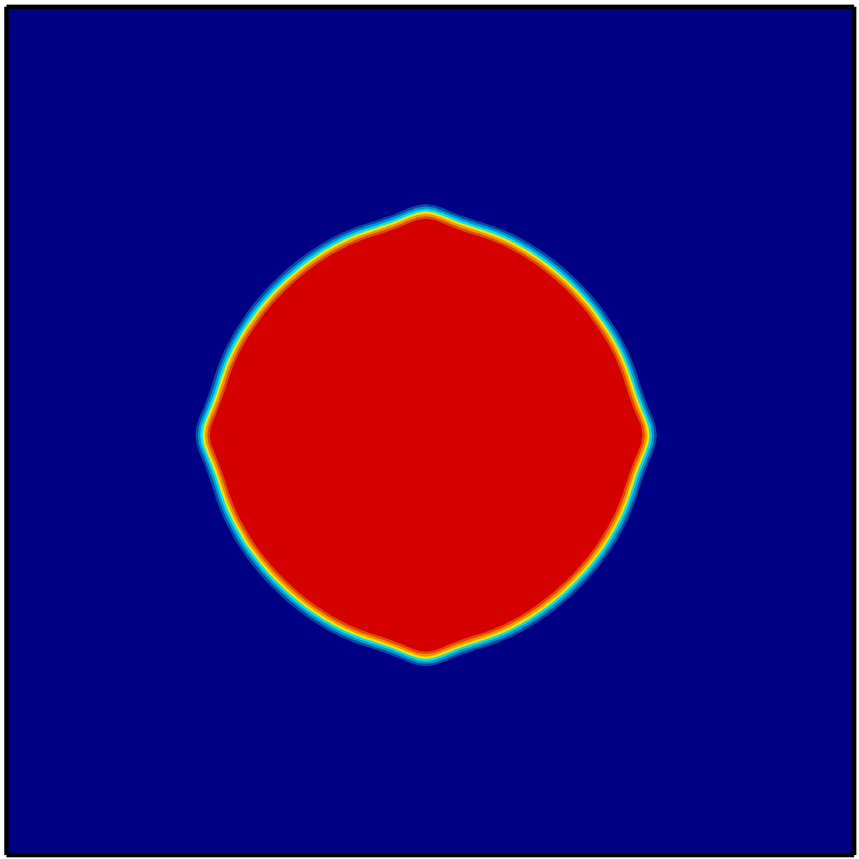}\quad\quad\includegraphics[width=0.2\textwidth]{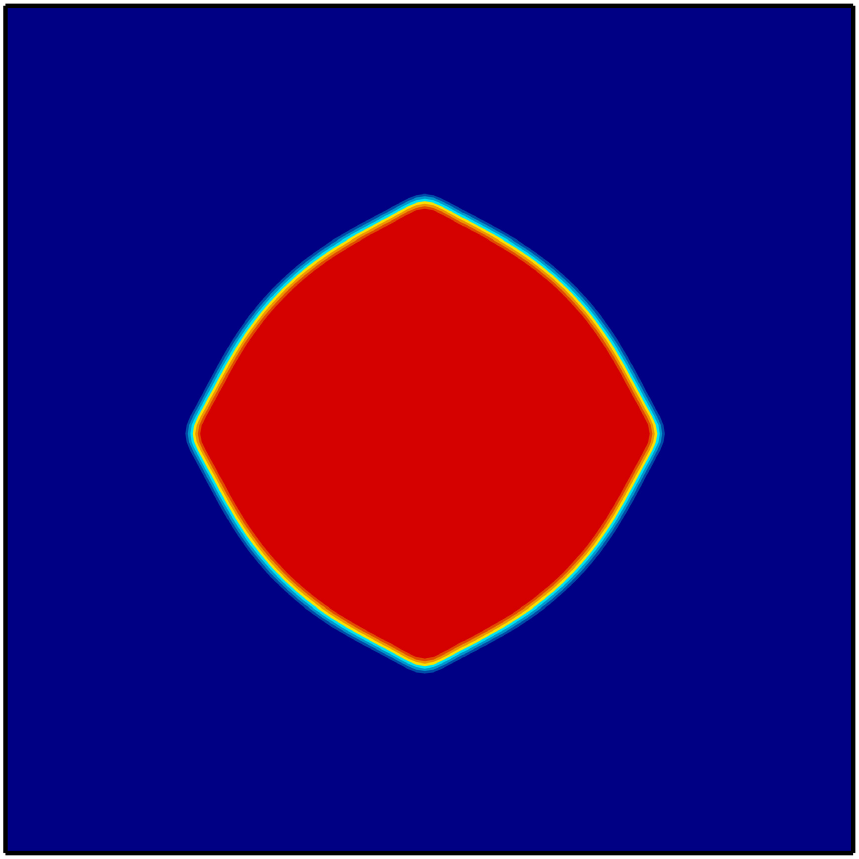}\quad\quad\includegraphics[width=0.2\textwidth]{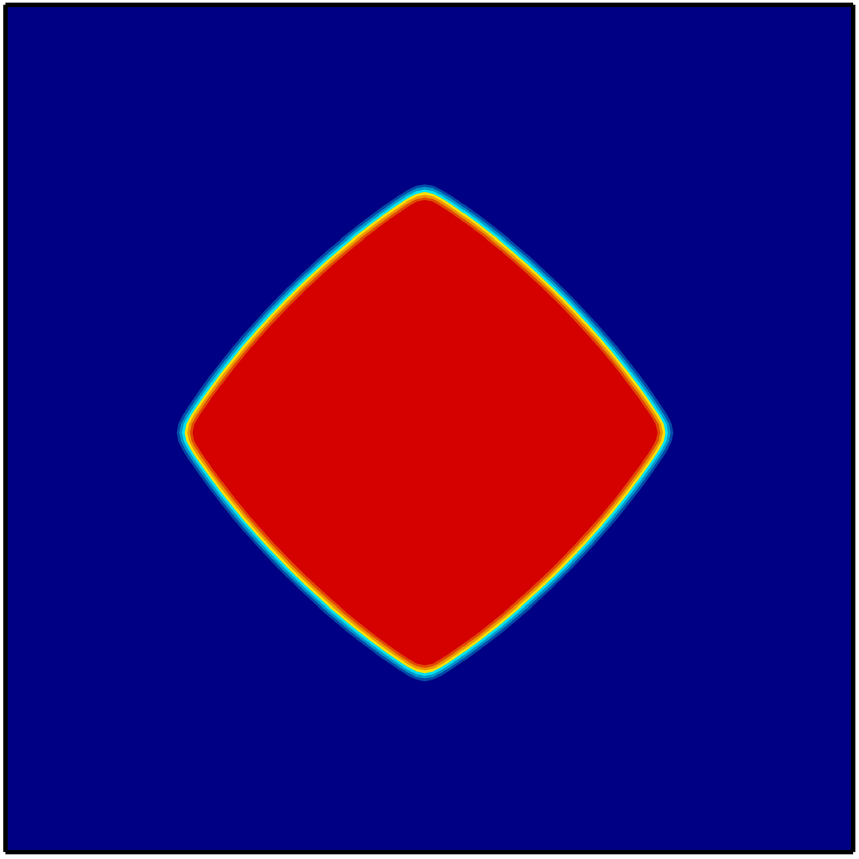}}
	\subfigure[$\alpha$=0.3]{\includegraphics[width=0.2\textwidth]{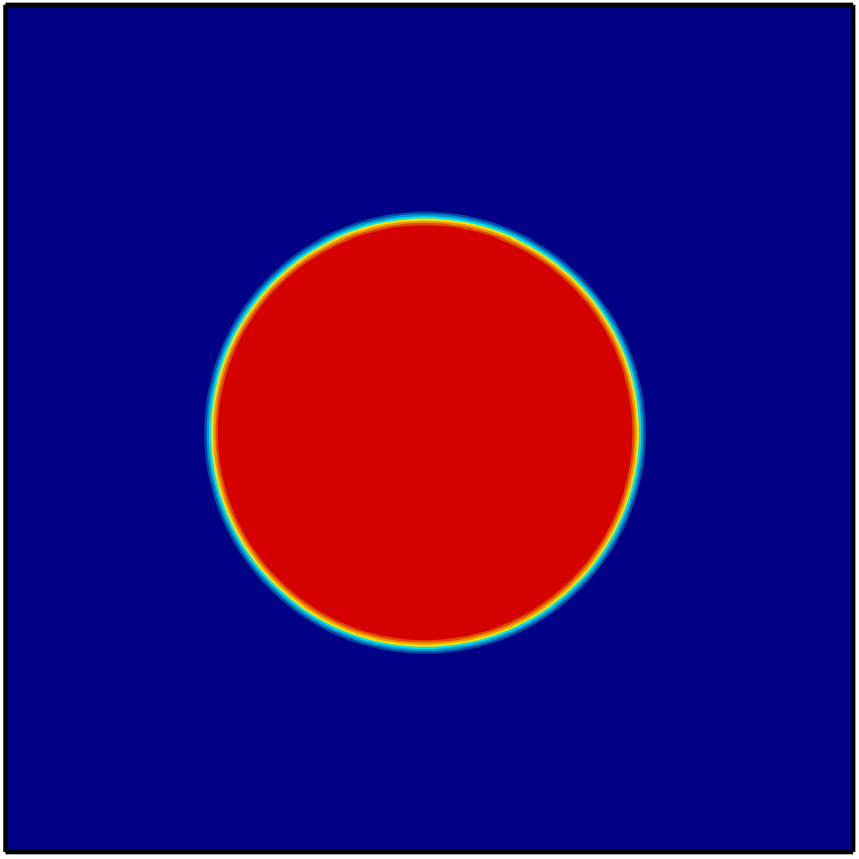}\quad\quad\includegraphics[width=0.2\textwidth]{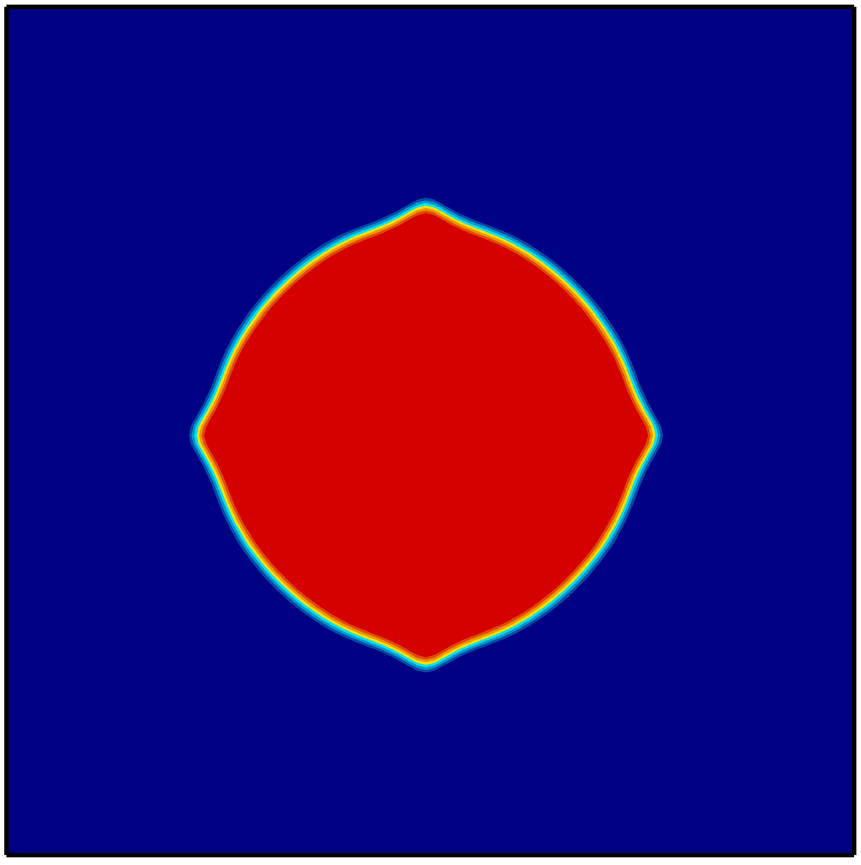}\quad\quad\includegraphics[width=0.2\textwidth]{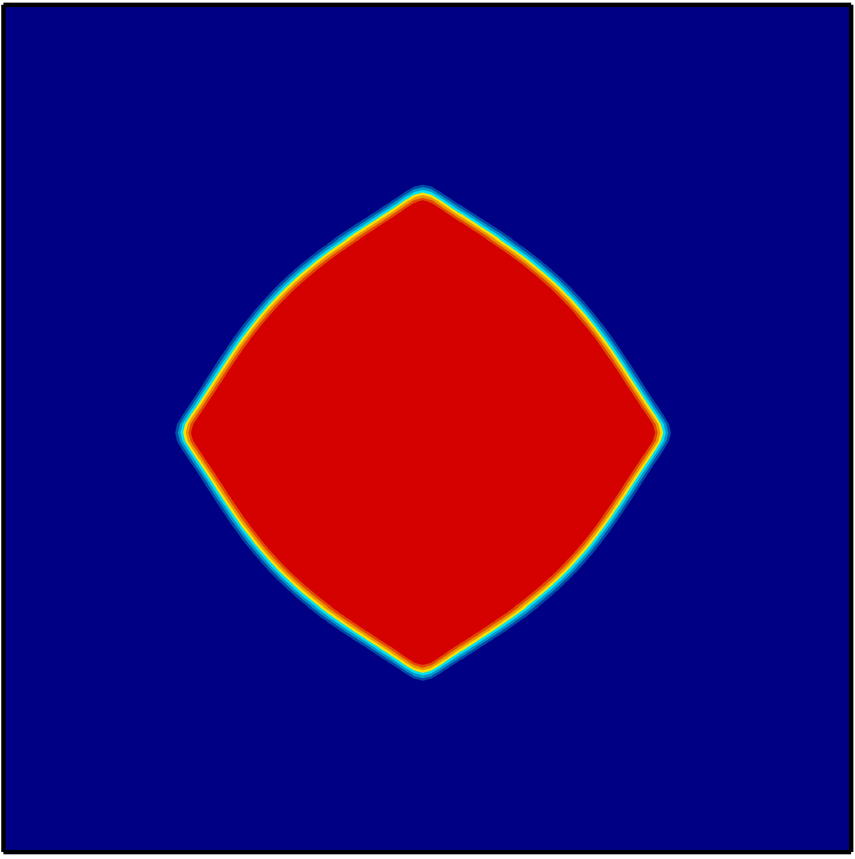}\quad\quad\includegraphics[width=0.2\textwidth]{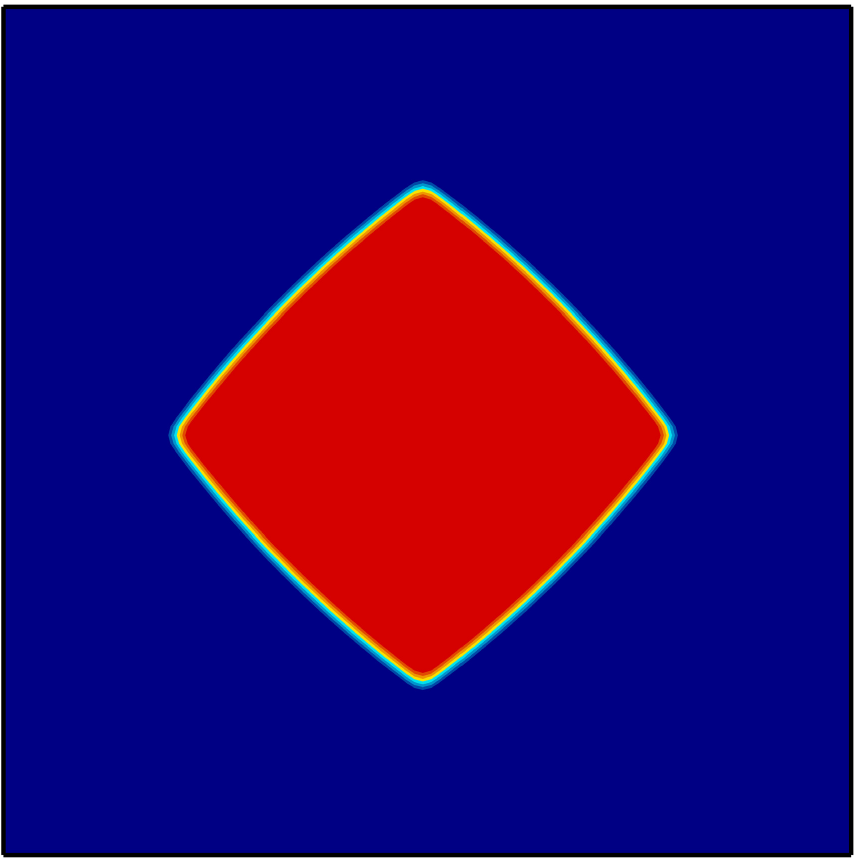}}
	\caption{The snapshots of the phase variable $\phi$ with four different time steps at $t=0$, $0.04$, $0.12$ and $2.0$ by using three anisotropic strength $\alpha$:(a) $\alpha=0.1$, (b) $\alpha=0.2$, (c) $\alpha=0.3$.}
	\label{fig2}
\end{figure}

In the first example, we consider the anisotropic evolution of a single droplet  located at the center of a computational domain. The initial condition is given by:
\begin{equation}
	\phi(x,y)=\tanh2\frac{R-\sqrt{(x-50)^{2}+(y-50)^{2}}}{D},
	\label{eq40}
\end{equation}
which enables its value to be smooth across the interface. In the simulation, The radius of the droplet is $R=25$ and the computational grid is chosen as  $100 \times 100$. To clearly demonstrate the impact of the anisotropic parameter $\alpha$, we perform calculations for various values of $\alpha$, specifically $\alpha=0.1,0.2,0,3$.  The other parameters are used from Eq. (\ref{eq39}). Fig. \ref{fig2} shows snapshots of the interfacial patterns for the three values of $\alpha$. Due to  the surface tension and anisotropic properties, the initial circular droplet evolves into a pyramid shape along four directions.   It can be seen that as $\alpha$ increases, we observe that the droplet will undergo a faster deformation.
When $\alpha=0.1$ is relatively small, there is not strong anisotropy enough. As $\alpha$ increases to $0.2$, the anisotropic effects become more pronounced. In particular, when $\alpha$  further increases larger to $0.3$, it leads to sharper anisotropic features at the four corners. 

\begin{figure}[H]
	\centering
	\subfigure{\includegraphics[width=0.5\textwidth]{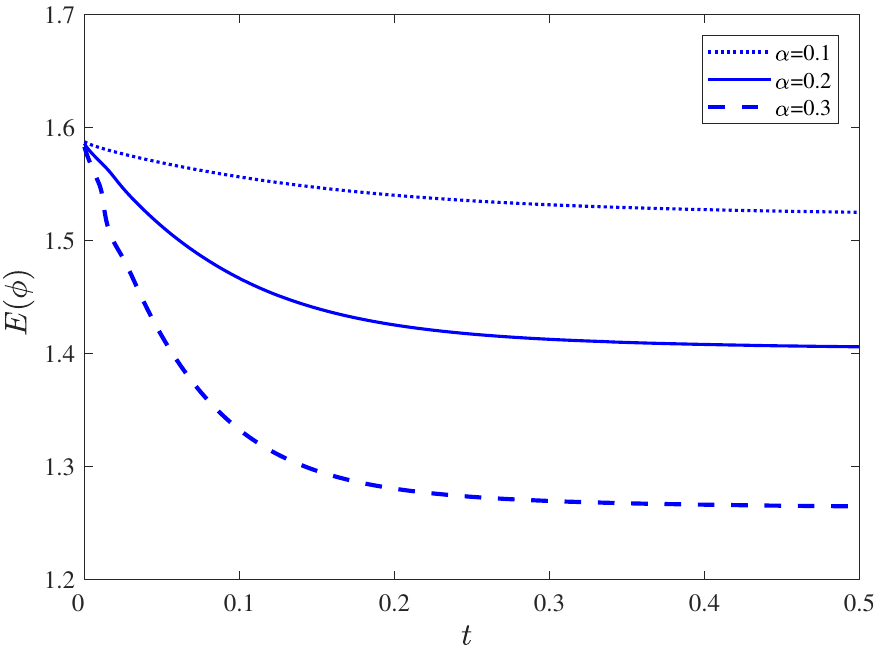}}
	
	\caption{Time evolution of the total free energy functional  with different anisotropic strength $\alpha$.}
	\label{fig3}
\end{figure}

In the classical phase
field theory, the total free energy tends to the minimum value with the time.  To give a quantitative analysis, Fig. \ref{fig3} gives the evolution of the free energy functional for each $\alpha$. we indeed observe that the energy at an arbitrary value of the anisotropic parameter $\alpha$ decreases with time.
As $\alpha$ gets larger, the energy decreases more rapidly. Besides, it is found that the energy stabilizes to a steady value for all cases. The present results fit well with simulations shown in \cite{Chen:cmame2019}.

\subsection{Anisotropic evolution   for two  droplets in 2D}
\begin{figure}[H]
	\centering
	\subfigure[]{\includegraphics[width=0.2\textwidth]{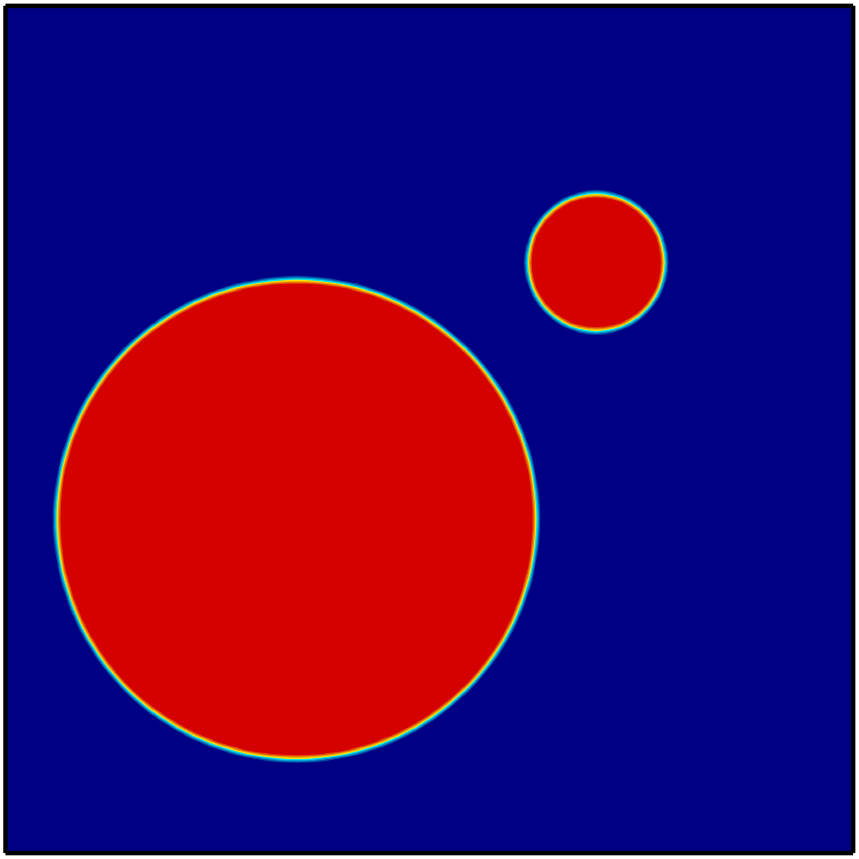}}\quad\quad
	\subfigure[]{\includegraphics[width=0.2\textwidth]{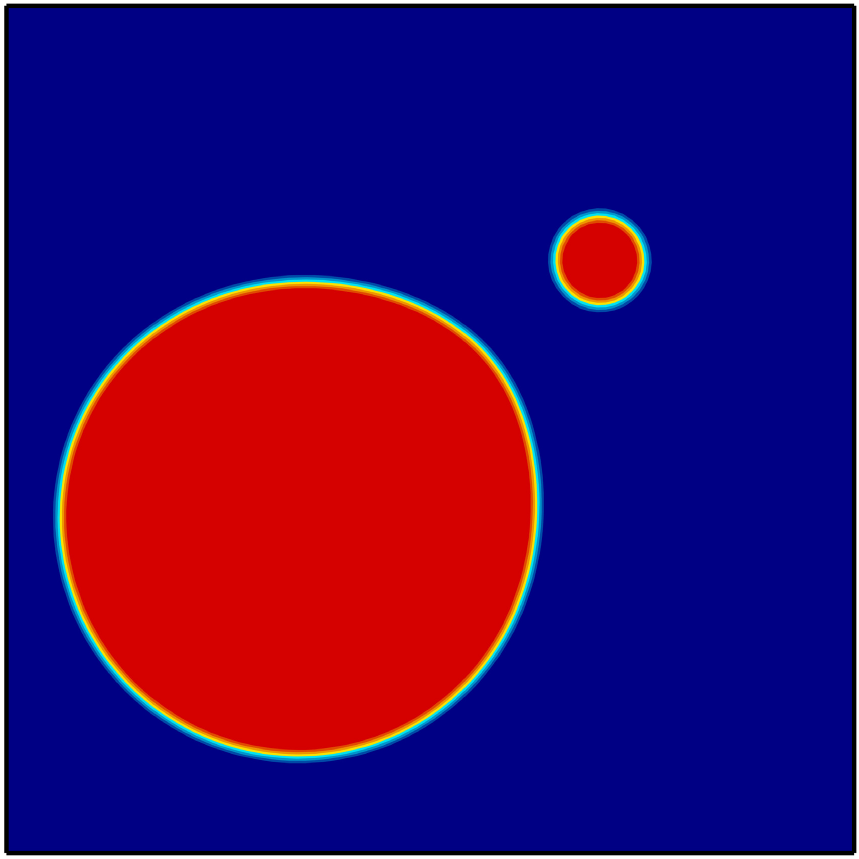}}\quad\quad
	\subfigure[]{\includegraphics[width=0.2\textwidth]{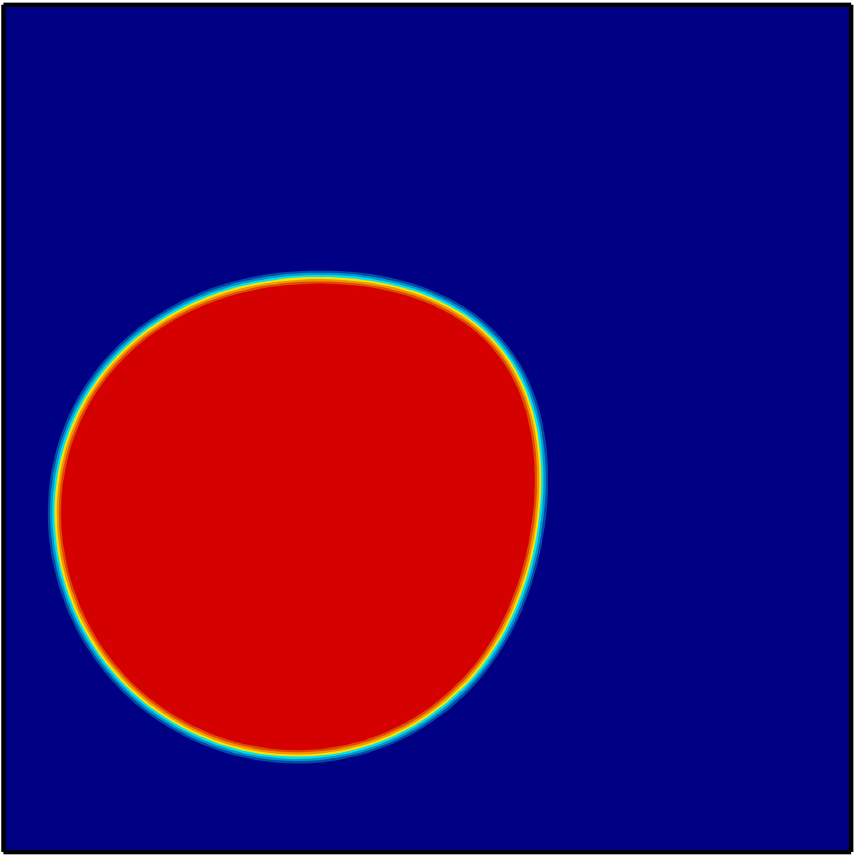}}\quad\quad
	\subfigure[]{\includegraphics[width=0.2\textwidth]{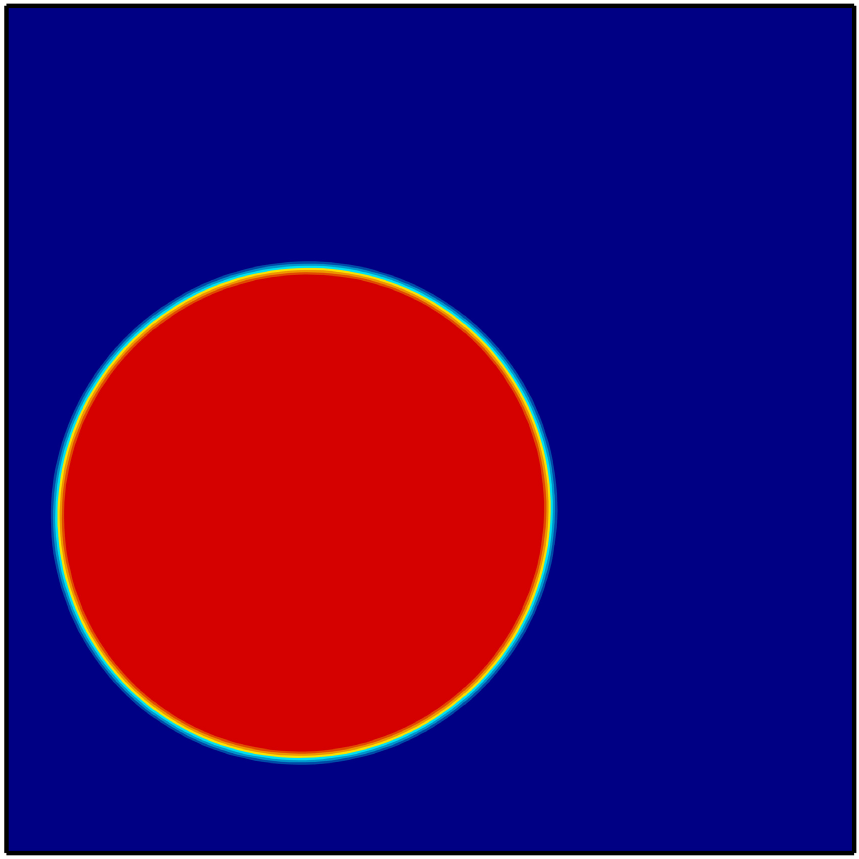}}
	\caption{The snapshots of the phase variable $\phi$  for the isotropic Cahn–Hilliard model with four different time steps $t=0$, $0.5$, $1.0$ and $8.0$}
	\label{fig4}
\end{figure}

\begin{figure}[H]
	\centering
	\subfigure[isotropic case]{\includegraphics[width=0.45\textwidth]{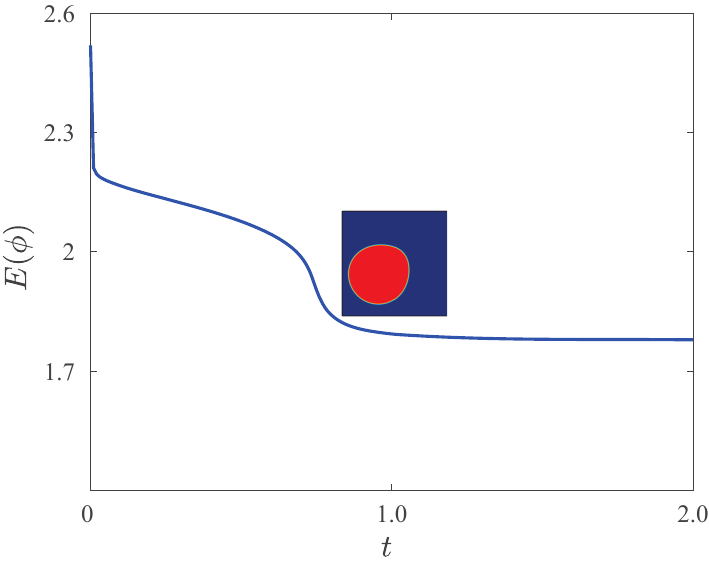}\label{fig5(a)}}
	\subfigure[anisotropic case]{\includegraphics[width=0.45\textwidth]{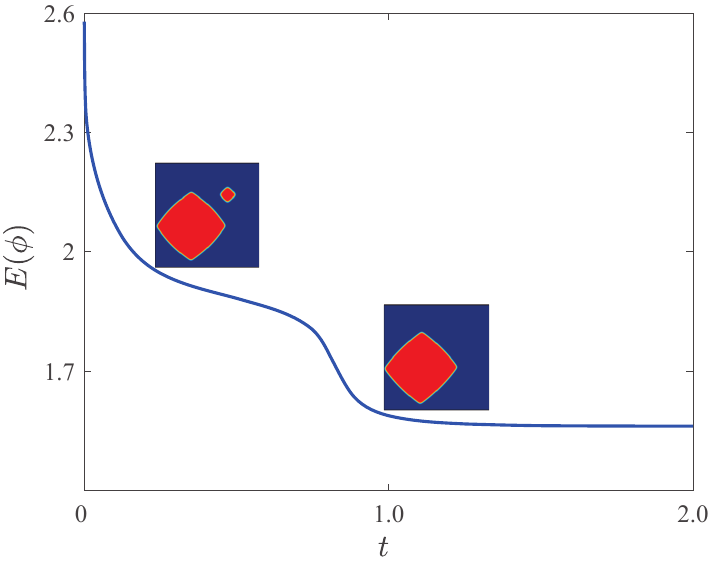}\label{fig5(b)}}
	\caption{Time evolution of the total free energy functional (a) isotropic case, (b) anisotropic case .}
	\label{fig5}
\end{figure}

In this example, to further validate the present LB method, we consider the evolution of two droplets using the same parameters (\ref{eq39}). This example has been previously tested by traditional numerical method \cite{Chen:cmame2019}. Initially, two different-sized circular droplets are located at the periodic region with $100 \times 100$ lattice cells. The distribution profile
of the order parameter is initialized by
\begin{equation}
	\phi(x, y)=\sum_{i=1}^2\tanh2\frac{R_i-\sqrt{(x-x_i)^2+(y-y_i)^2}}{D})+1,
	\label{eq41}
\end{equation}
where the centers and radii of the droplets are $(R_1,x_1,y_1)=(28,35,40)$ and $(R_2,x_2,y_2)=(9,70,70)$, respectively. Firstly, a special case of the  isotropic  Cahn–Hilliard system where $\gamma(\boldsymbol{n})=1$ is considered. 
Figs.  \ref{fig4} show the zero-isocontours of $\phi$ at various moments. From  Figs. \ref{fig4}, we can observe that the small droplet gradually is absorbed by the big droplet as time changes.  Additionally, we also 
present the energy evolution using the energy functional equation (\ref{eq7}) in Fig. \ref{fig5(a)}. It can be clearly seen that the energy decreases gradually, and once the absorption is complete, the energy  drop rapidly and the system reaches equilibrium.

\begin{figure}[H]
	\centering
	\subfigure[]{\includegraphics[width=0.2\textwidth]{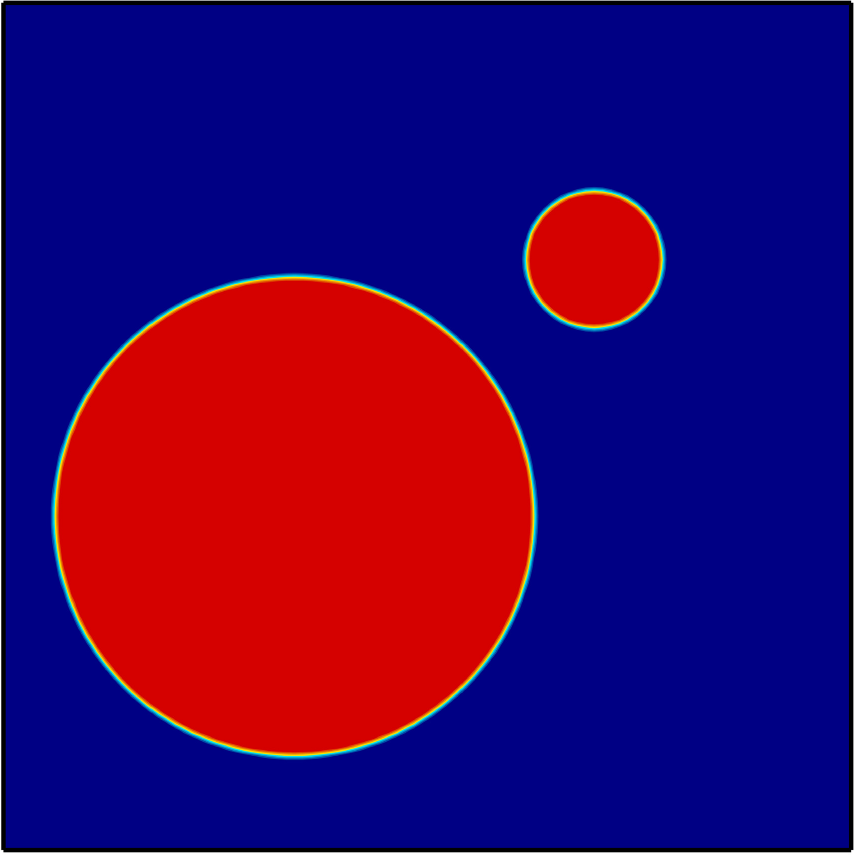}}\quad\quad
	\subfigure[]{\includegraphics[width=0.2\textwidth]{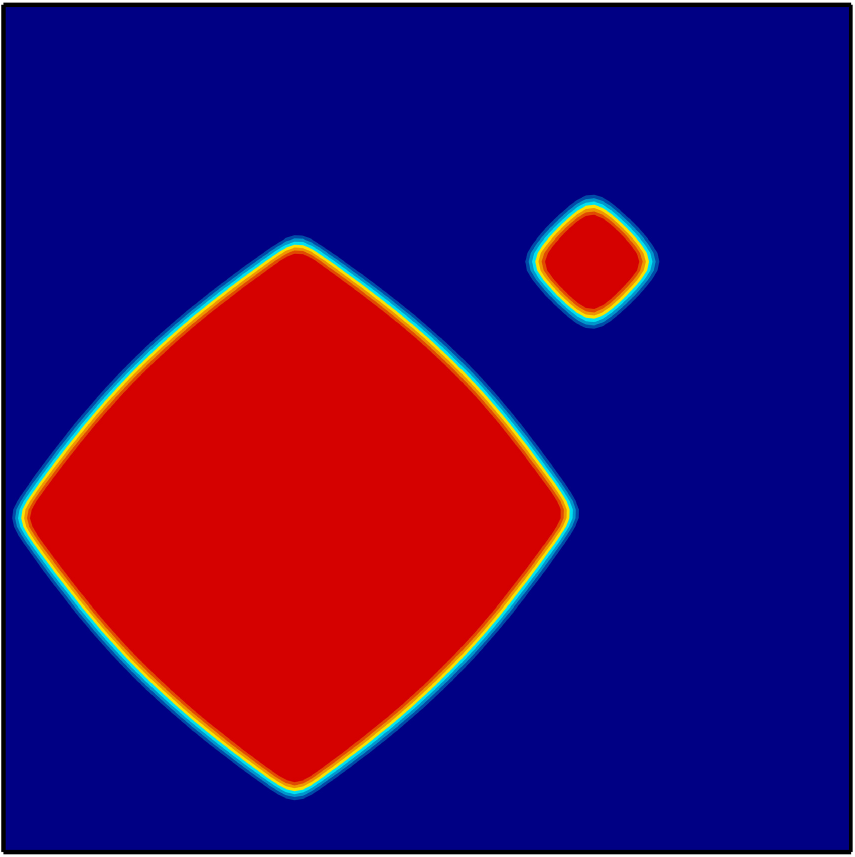}}\quad\quad
	\subfigure[]{\includegraphics[width=0.2\textwidth]{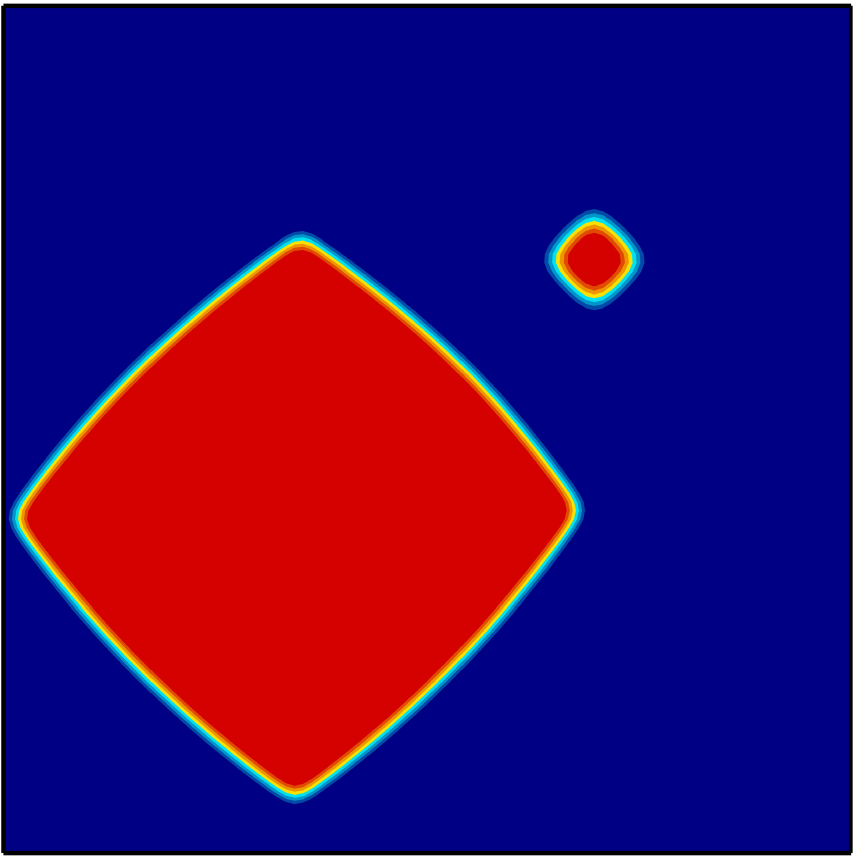}}\quad\quad
	\subfigure[]{\includegraphics[width=0.2\textwidth]{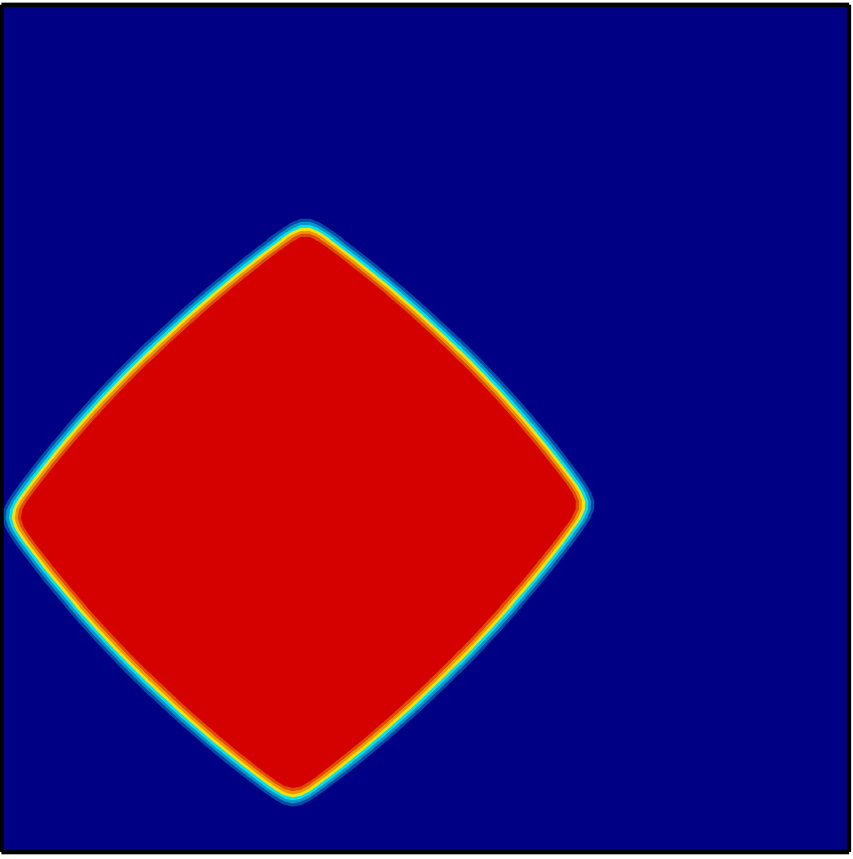}}
	\caption{The snapshots of the phase variable $\phi$  for the  anisotropic Cahn–Hilliard model at $\phi=0.3$ with four different time steps $t=0$, $0.4$, $0.7$ and $2.0$}
	\label{fig6}
\end{figure}

Next, We  consider the evolution of two droplets using the parameters specified in Eqs. (\ref{eq39}) and initial condition (\ref{eq41}), but the anisotropic parameter $\alpha$ is set to be $0.3$. Fig. \ref{fig6} displays snapshots of the profiles of the phase field variable $\phi$ at differnt time. Similar to the isotropic case, the smaller droplet is fully absorbed by the larger one. However, during the absorption process, the droplets first exhibit anisotropic behavior. As seen in  Fig. \ref{fig5(b)}, the energy undergoes two rapid declines: the first occurs when the droplets form pyramid shapes at the four corners, and the second when the smaller droplet disappears. Finally, the energy gradually stabilizes. Comparing the energy changes in Figs. \ref{fig5(a)} and \ref{fig5(b)}, we can see that the energy decreases over time and eventually stabilizes at a constant value. The inclusion of anisotropic parameter $\alpha$ clearly accelerates this decrease.

\subsection{Anisotropic spinodal decomposition in 2D}

\begin{figure}[H]
	\centering
	\subfigure[]{\includegraphics[width=0.2\textwidth]{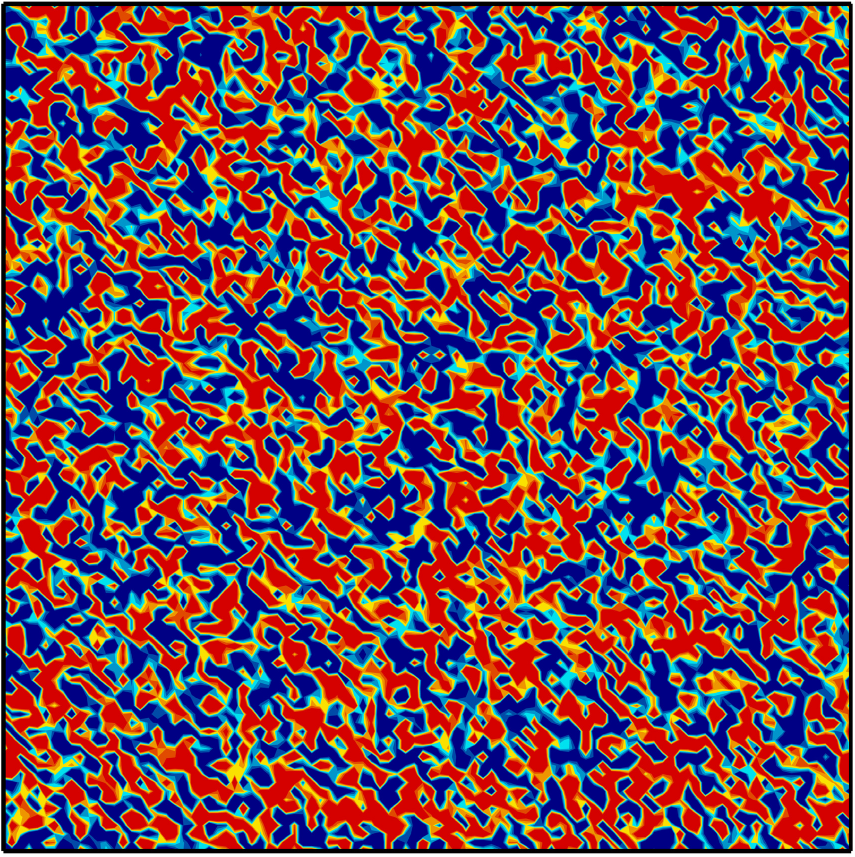}}\quad\quad
	\subfigure[]{\includegraphics[width=0.2\textwidth]{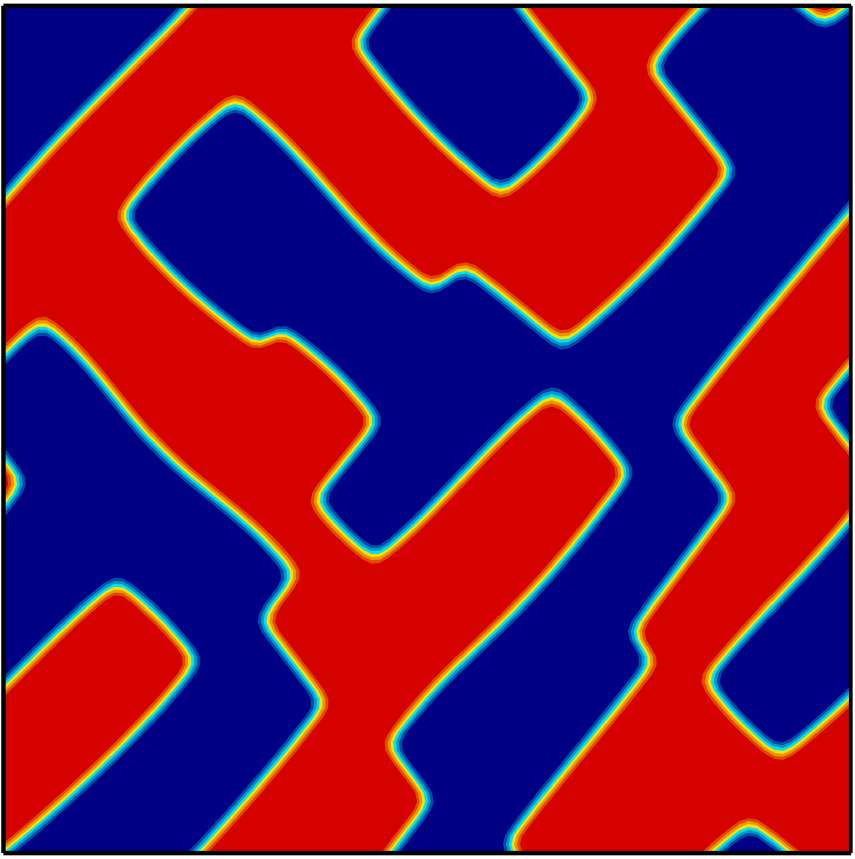}}\quad\quad
	\subfigure[]{\includegraphics[width=0.2\textwidth]{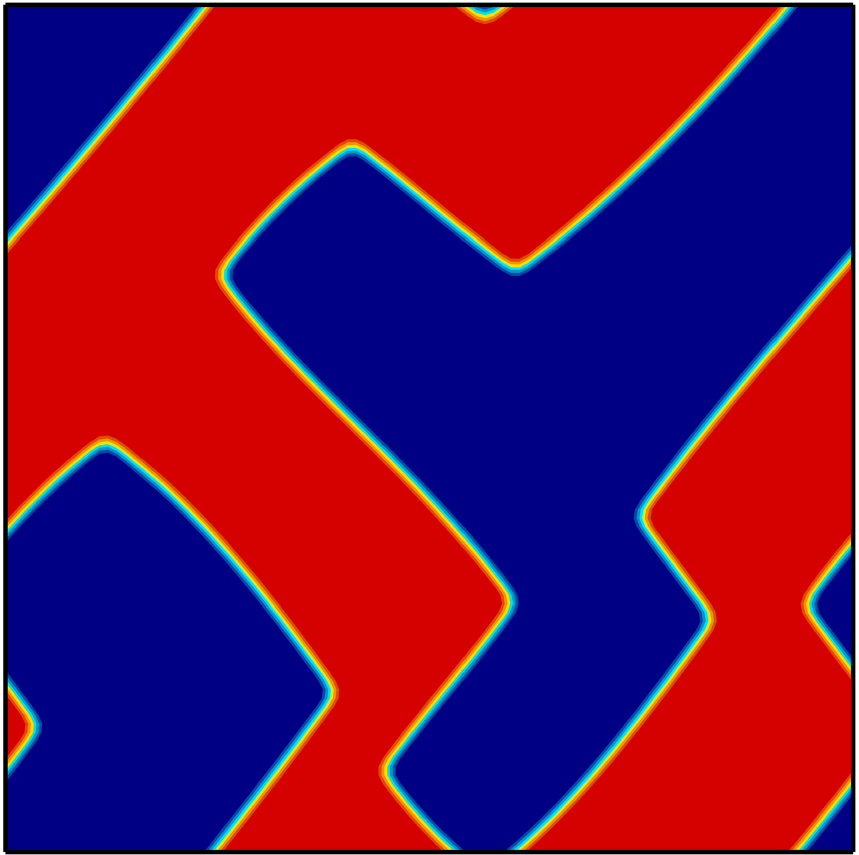}}\\
	\subfigure[]{\includegraphics[width=0.2\textwidth]{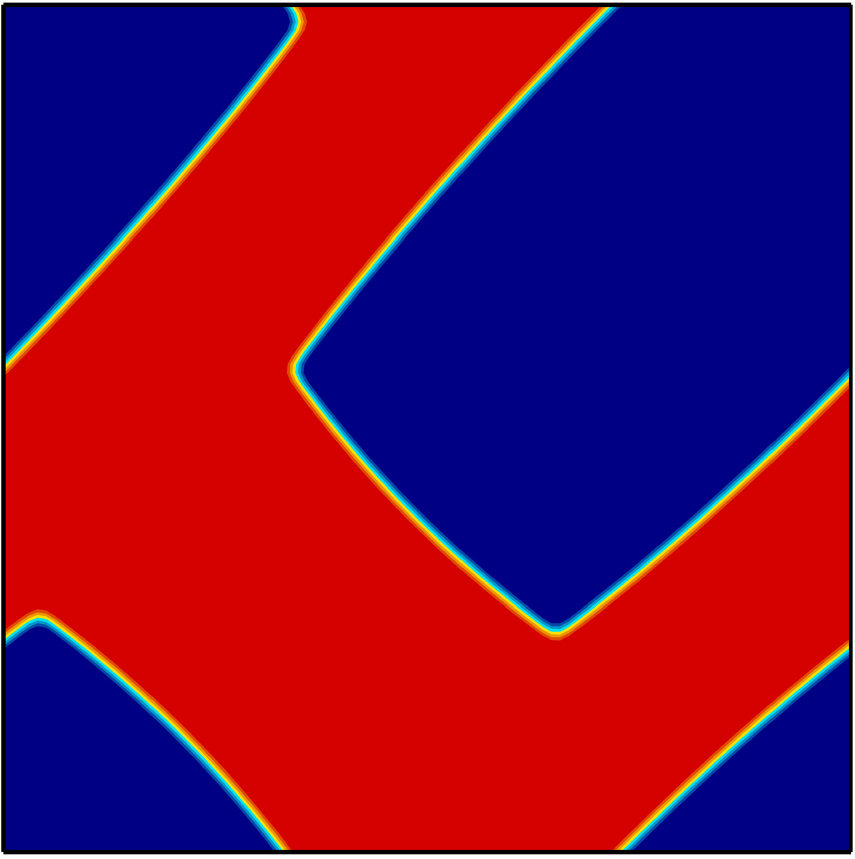}}\quad\quad
	\subfigure[]{\includegraphics[width=0.2\textwidth]{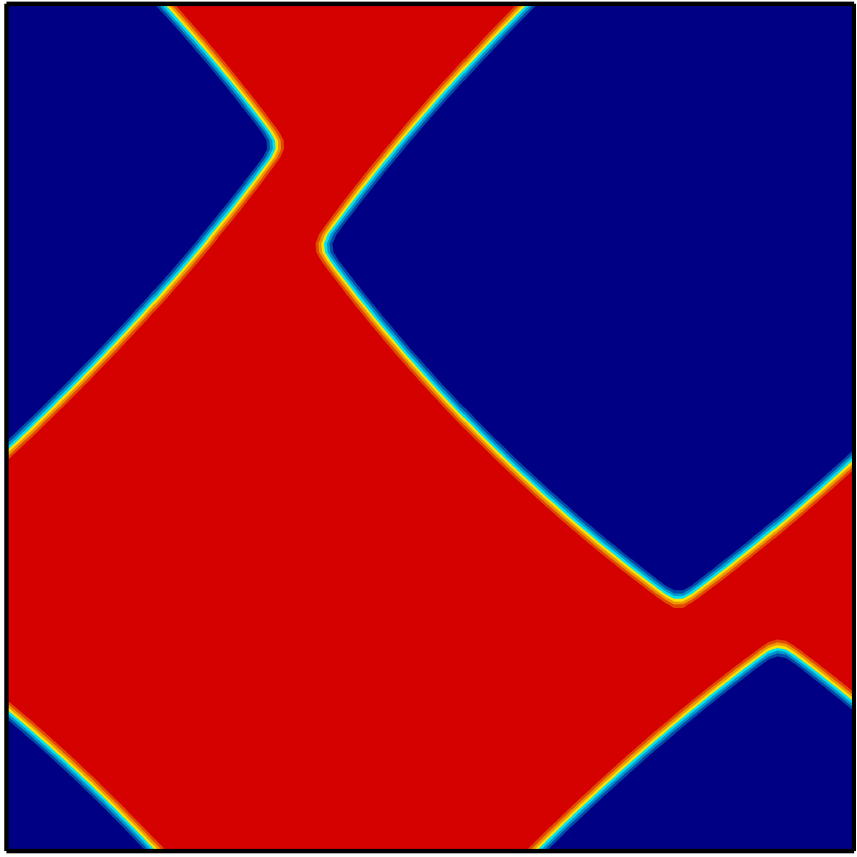}}\quad\quad
	\subfigure[]{\includegraphics[width=0.2\textwidth]{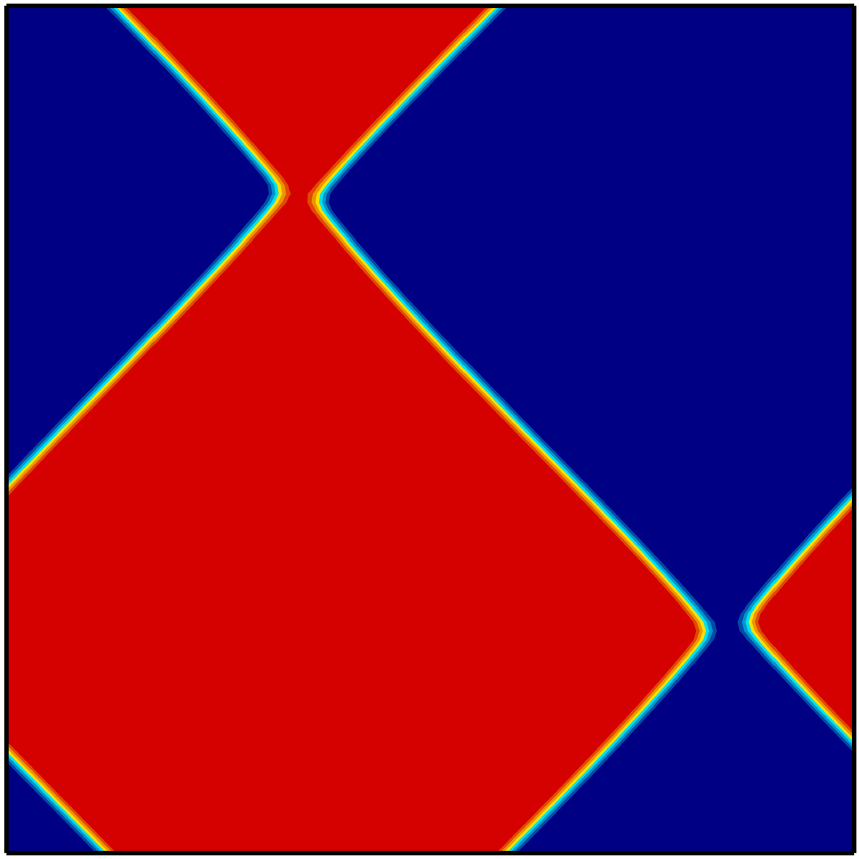}}

	\caption{The snapshots of the phase variable $\phi$  for the spinodal decomposition example with six different time steps $t=0$, $1.0$, $5.0$, $30$, $70$ and $90$.}
	\label{fig7}
\end{figure}

\begin{figure}[H]
	\centering
	\subfigure{\includegraphics[width=0.5\textwidth]{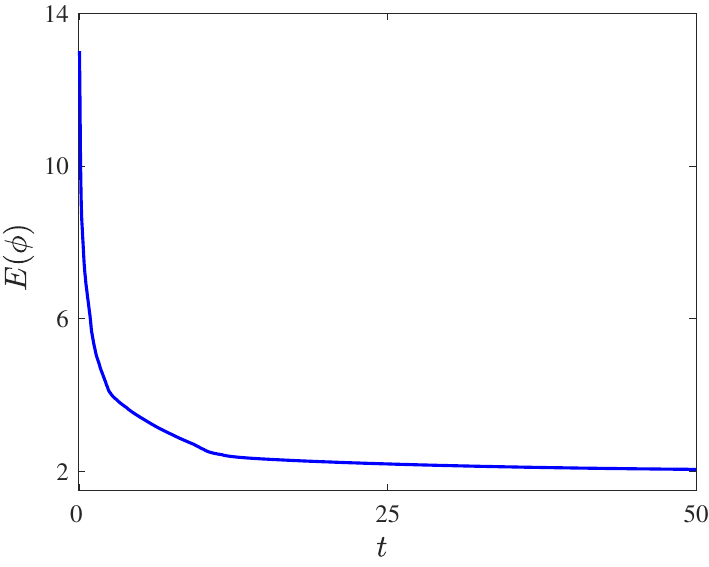}}
	
	\caption{Time evolution of the total free energy functional for the spinodal decomposition example.}
	\label{fig8}
\end{figure}

Additionally, we consider a case with a more randomized initial condition.  The spinodal decomposition is an interesting phenomenon, which has been extensively studied by phase-field model, with few works reported in the anisotropic case. Therefore, we simulate this complex  dynamic problem of anisotropic phase separation to validate  the current LB method. The initial interface profile in the computational domain of $100 \times 100$ can be expressed by 
\begin{equation}
	\phi(x,y)=\mathrm{rand}(x,y),
		\label{eq42}
\end{equation}
where $\mathrm{rand}(x,y)$ is a random function that generates values between $-1$ and $1$. The anisotropic parameter $\alpha$ is also set to be $0.3$. In this case, the spinodal decomposition  occurs, leading to the formation of a two-phase state driven by unstable concentration fluctuations. From Fig. \ref{fig7} ,we can find that these small perturbations grow with time and then several small anisotropic droplets are formed. These droplets will gradually increase in size and coalesce into larger droplets which will eventually reach a stable and sharp pyramid shape. Furthermore, we plot the  evolution of the total energy in Fig. \ref{fig8}, which is decreasing over time and consistent to previous work \cite{Chen:cmame2019}.

\subsection{Anisotropic evolution for droplets in 3D  }
\begin{figure}[H]
	\centering
	\subfigure[]{\includegraphics[width=0.2\textwidth]{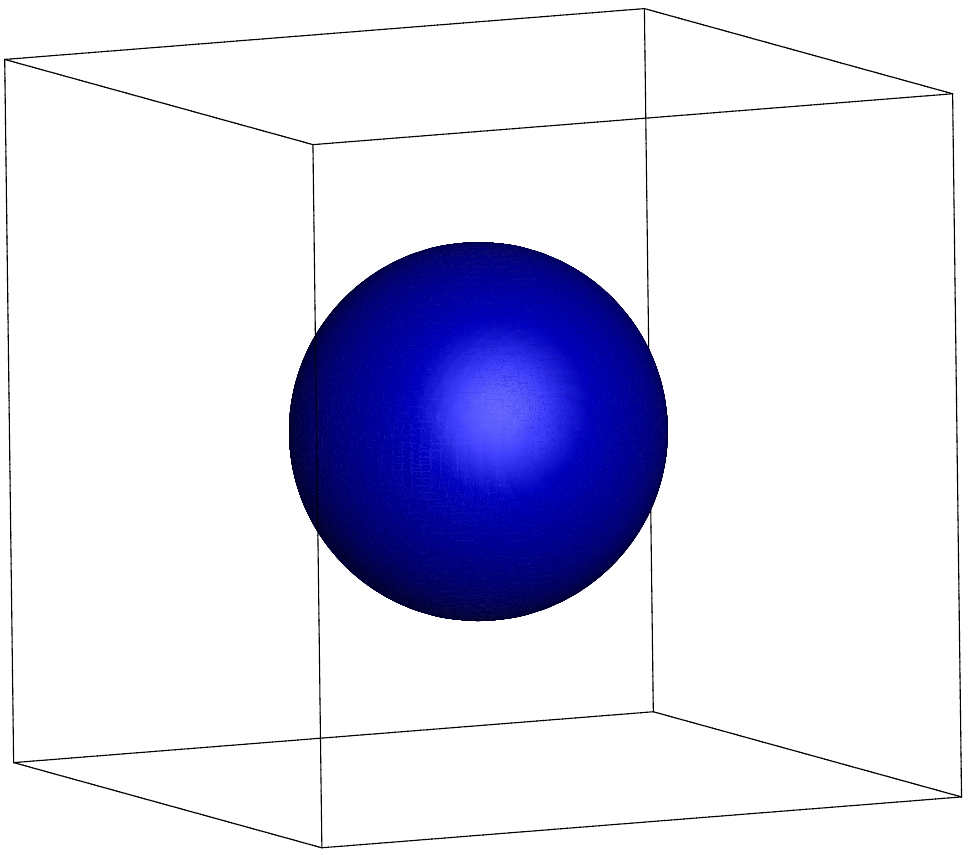}}\quad\quad
	\subfigure[]{\includegraphics[width=0.2\textwidth]{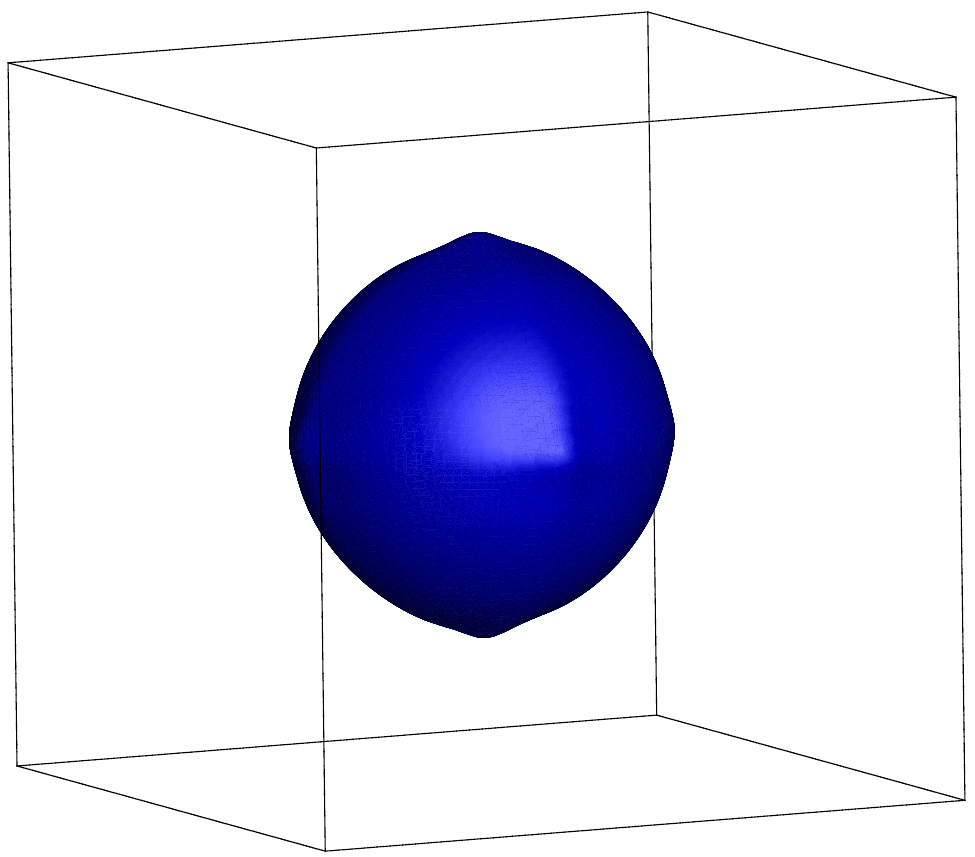}}\quad\quad
	\subfigure[]{\includegraphics[width=0.2\textwidth]{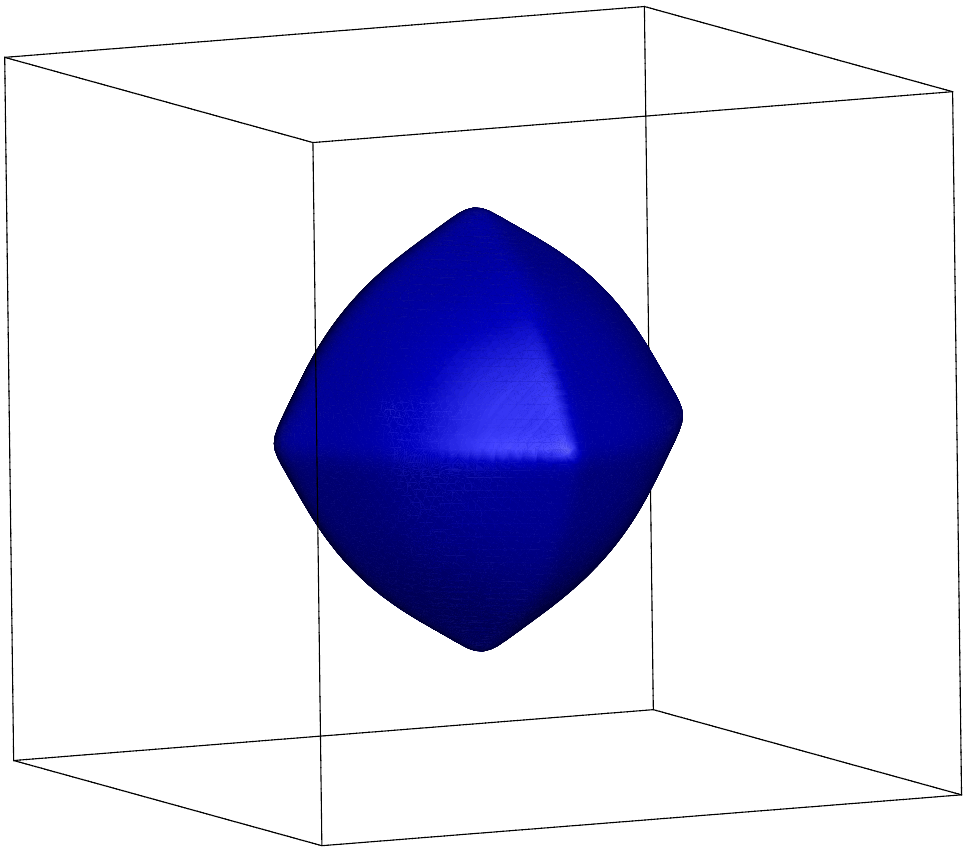}}\quad\quad
	\subfigure[]{\includegraphics[width=0.2\textwidth]{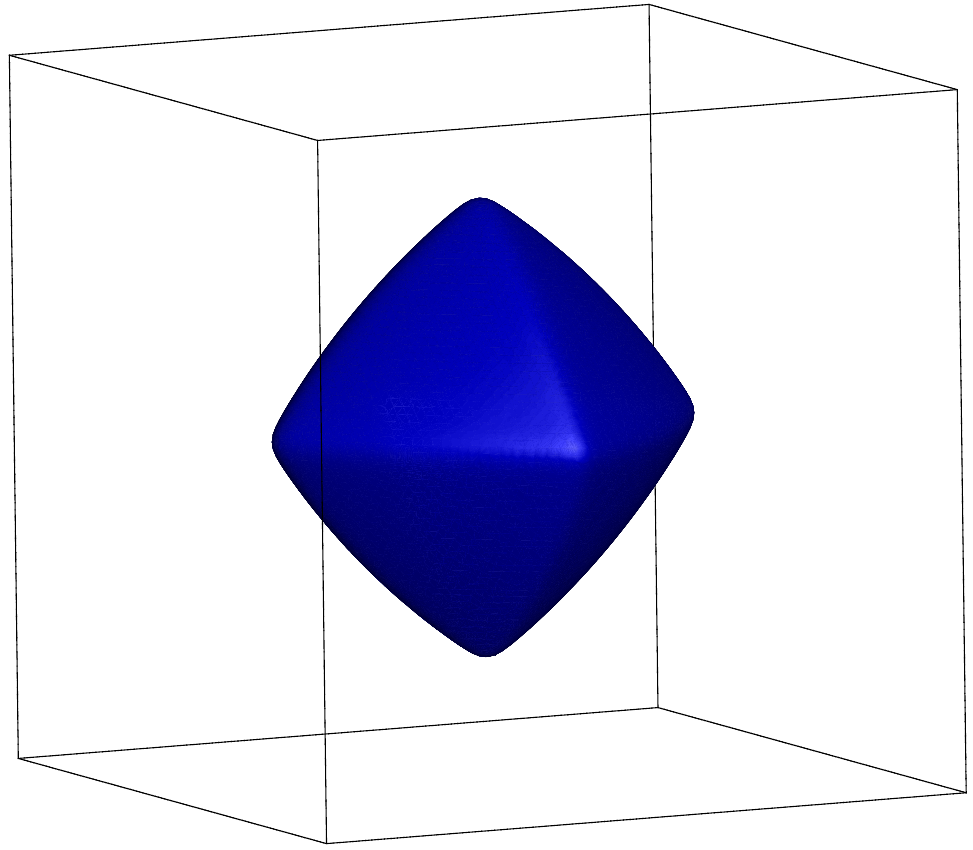}}
	\caption{The snapshots of the phase variable $\phi$  for  the anisotropic model of the 3D droplrt with four different time steps $t=0$, $0.02$, $0.1$ and  $0.7$.}
	\label{fig9}
\end{figure}

\begin{figure}[H]
	\centering
	\subfigure{\includegraphics[width=0.5\textwidth]{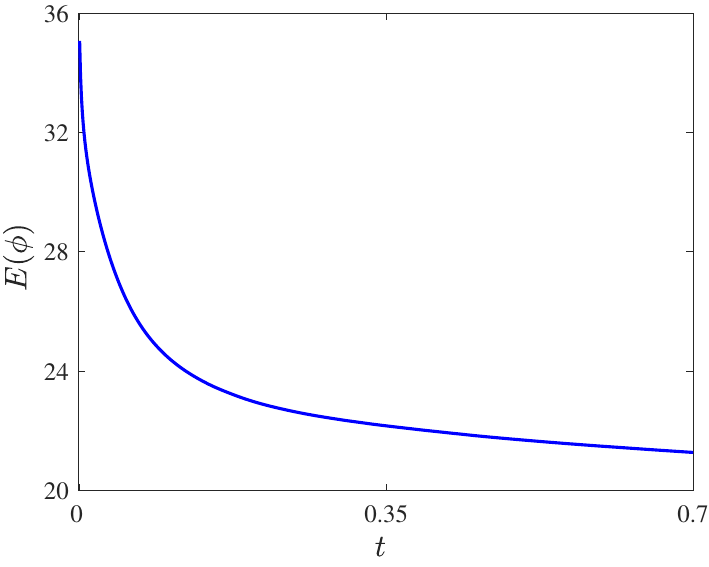}}
	
	\caption{Time evolution of the total free energy functional for  the anisotropic model of the 3D droplrt.}
	\label{fig10}
\end{figure}

All the above simulations on the evolution of droplets are based on the 2D cases. In order to show the present LB method is capable of studying anisotropic evolution in 3D case, the 3D droplets are simulated in this subsection.
To begin with, we consider a relatively simple case  involving the anisotropic evolution of a single droplet.  The initial configuration of the order parameter distribution is defined as:
\begin{equation}
	\phi(x,y,z)=\tanh2\frac{R-\sqrt{(x-50)^2+(y-50)^2+(z-50)^2}}{D}.
		\label{eq43}
\end{equation}
Here, $R=25$ represents droplet radius. Some used physical parameters in the simulation are fixed by Eqs. (\ref{eq39}). Figs. \ref{fig9} depict the time evolution of the interfacial patterns. Under the influence of surface tension and anisotropy, the 3D droplet begins to contract, leading to interface deformation, and eventually evolves into an anisotropic pyramid with missing orientations at the six corners. Further, Fig. \ref{fig10} shows  present the evolution of the free energy functional until the system reaches a steady state. These results are in excellent agreement with those reported in \cite{Torabi:mpes2009,Chen:cmame2019}.

\begin{figure}[H]
	\centering
	\subfigure[]{\includegraphics[width=0.2\textwidth]{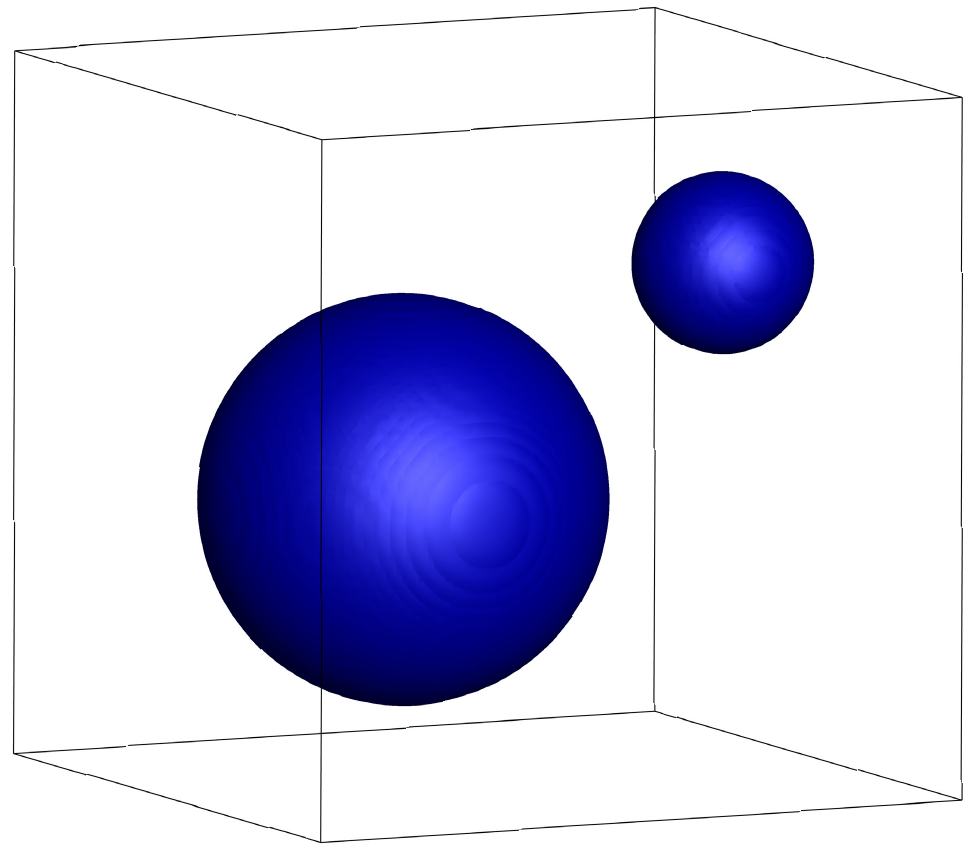}}\quad\quad
	\subfigure[]{\includegraphics[width=0.2\textwidth]{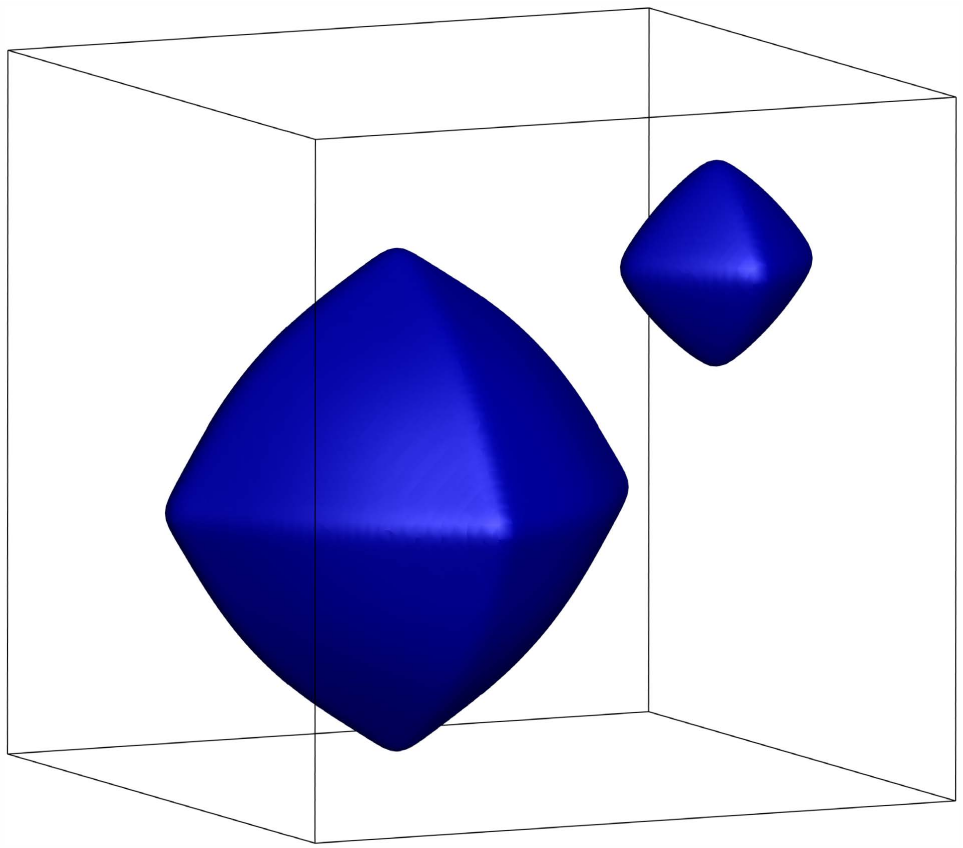}}\quad\quad
	\subfigure[]{\includegraphics[width=0.2\textwidth]{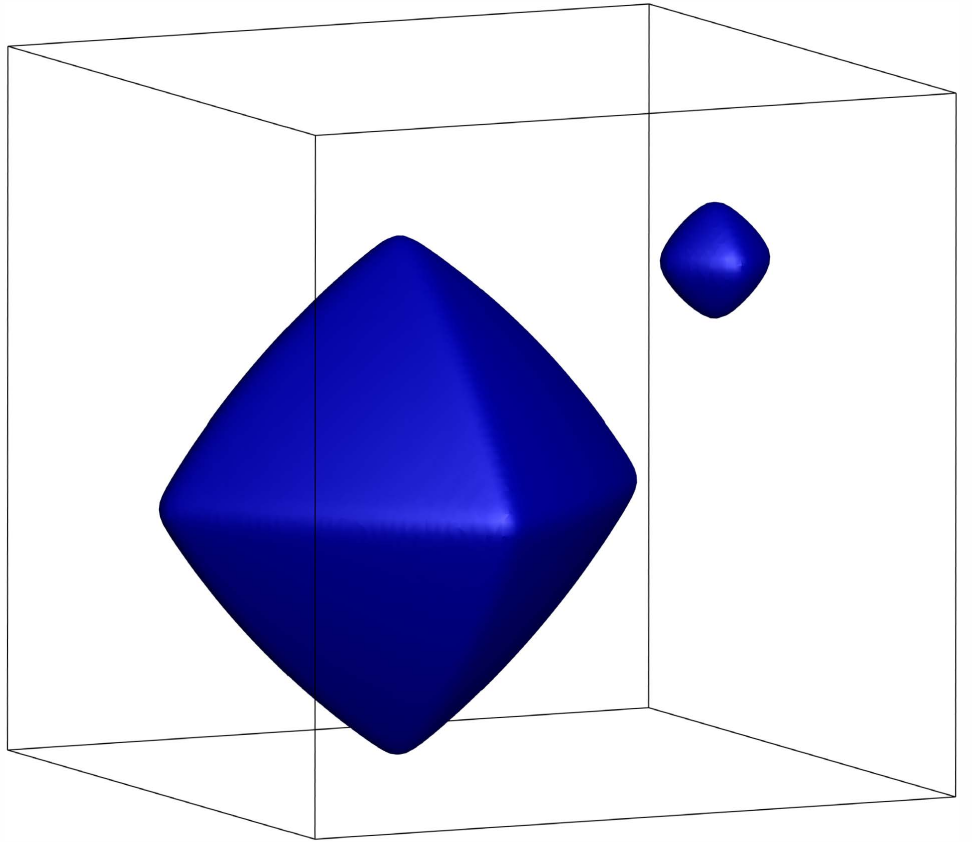}}\quad\quad
	\subfigure[]{\includegraphics[width=0.2\textwidth]{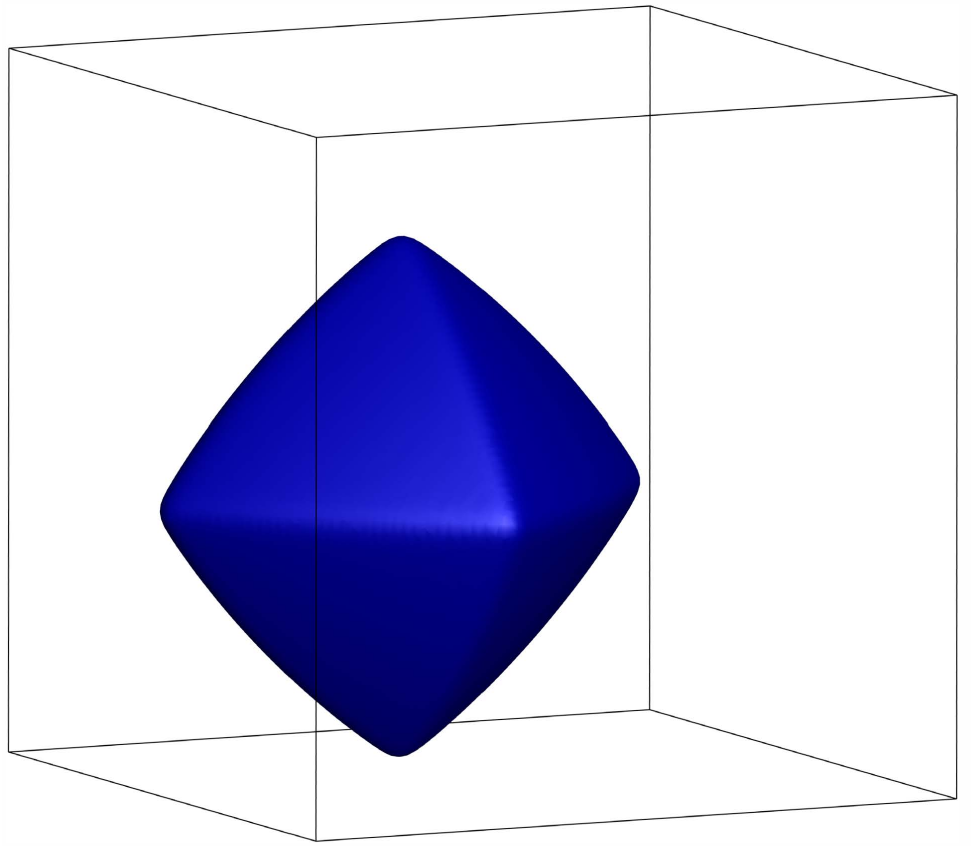}}
	\caption{The snapshots of the phase variable $\phi$  for  the anisotropic model of two 3D droplrts with four different time steps $t=0$, $0.2$, $0.8$ and $1.5$.}
	\label{fig11}
\end{figure}

\begin{figure}[H]
	\centering
	\subfigure[]{\includegraphics[width=0.5\textwidth]{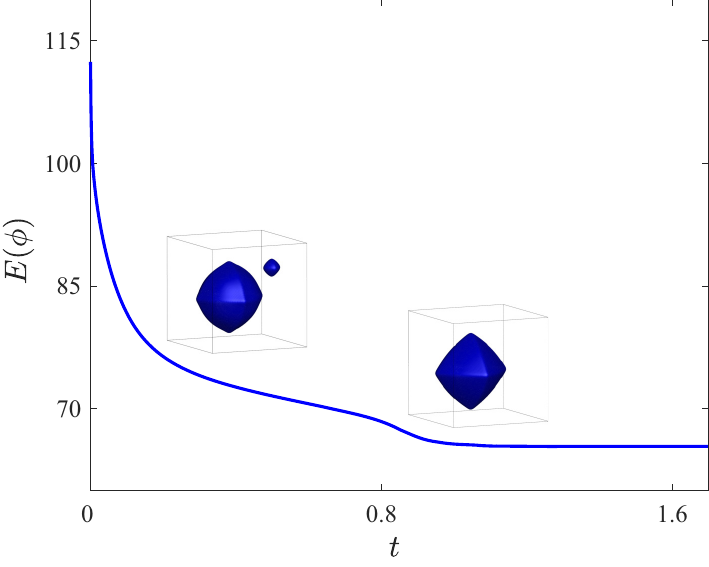}}
	
	\caption{Time evolution of the total free energy functional for  the anisotropic model of two 3D droplrts.}
		\label{fig12}
\end{figure}

Finally, to show the superiority of the present LB method in modelling anisotropic evolution in 3D cases, we perfrom the evolution of two droplets by using the following initial condition
\begin{equation}
	\phi(x,y,z)=\sum_{i=1}^2\tanh2\frac{R_i-\sqrt{(x-x_i)^2+(y-y_i)^2+(z-z_i)}}{D}+1,
\end{equation}
where $(R_1,x_1,y_1,z_1)=(28,45,40,40)$ and $(R_2,x_2,y_2,z_1)=(12,75,75,75)$. In this test, the computational grid is chosen as $100 \times 100\times 100$, Figs. \ref{fig11} illustrate that, similar to the 2D example in Figs. \ref{fig6}, the two spheres initially transform into anisotropic pyramids with missing orientations at the six corners. As the anisotropic system evolves, the smaller pyramid is eventually absorbed by the larger one. Additionally, Fig. \ref{fig12} shows that the free energy function decreases as it evolves toward a steady state, with two distinct rapid reductions observed during the process.  Our results  fit well with
previous works \cite{Chen:cmame2019}.

\section{Conclusion}
The anisotropic Cahn-Hilliard equation is a classical diffusive interface model often used to simulate the formation of pyramids on nanoscale crystal surfaces; however, numerical modeling of this equation remains a significant challenge in the LB community. In this paper, we propose an MRT-LB method pecifically designed for anisotropic CH equations.  Through a series of algebraic manipulations, the anisotropic CH equation is reformulated as a standard convection-diffusion equation with source terms. The MRT-LB model is then constructed based on this nonlinear equation, with the MRT model enhancing the stability of the numerical calculations. Using CE analysis, it can accurately recover the macroscopic governing equation. The validity of the proposed LB method is demonstrated through numerical simulations of anisotropic droplet evolution, droplet absorption, and spinodal decomposition in both 2D and 3D cases.The results show  that the model accurately simulates the evolution of anisotropy, with the calculated  system energy conforming to the laws of energy dissipation. Additionally, the  predictions of model are consistent with previously reported results, confirming its suitability for the simulation of anisotropic CH equation.

\section*{Acknowlgedgments }

This work is financially supported by the National Natural Science Foundation of China (Grant No.12472297).

\newpage	




\begin{thebibliography}{1}
	
	\bibitem{Mouketou:cppm2019} 
	Mouketou FN, Kolesnikov A. Modelling and simulation of multiphase flow applicable to processes in oil and gas industry. Chem. Prod. Process Model. 2019;14(1):20170066.
	
	
	\bibitem{Tokeshi:ac2002} 
	Tokeshi M, Minagawa T, Uchiyama K, Hibara A, Sato K, Hisamoto H, Kitamori T. Continuous-flow chemical processing on a microchip by combining microunit operations and a multiphase flow network. Anal. Chem. 2002;74(7):1565-1571.
	
	
	\bibitem{Jiang:pecs2010} 
	Jiang X, Siamas GA, Jagus K, Karayiannis TG. Physical modelling and advanced simulations of gas–liquid two-phase jet flows in atomization and sprays.  Prog Energy Combust Sci 2010;36(2):131-167.
	
	\bibitem{Dudukovic:etfs2002} 
	Dudukovic MP. Opaque multiphase flows: experiments and modeling. Exp. Therm. Fluid 
	Sci. 2002;26(6-7):747-761.
	
	
	\bibitem{Yating:ciesc2024} 
	Yating LI, Zhongdong W, Yanpeng D, Chunying Z, Youguang M,  Taotao F. Research progress of capillary flow in microchannels and its engineering application. CIESC Journal, 2024;75(1):159.
	
	
	
	\bibitem{Sussman:jcp1994} 
	Sussman M, Smereka P, Osher S. A level set approach for computing solutions to incompressible two-phase flow. J Comput Phys 1994;114(1):146-159.
	
	\bibitem{Hirt:jcp1981} 
	Hirt CW, Nichols BD. Volume of fluid (VOF) method for the dynamics of free boundaries. J. Comput. Phys. 1981;39(1):201-225.
	
	\bibitem{Raeini:jcp2012} 
	Raeini AQ, Blunt MJ, Bijeljic B. Modelling two-phase flow in porous media at the pore scale using the volume-of-fluid method. J. Comput. Phys. 2012;231(17):5653-5668.
	
	
	
	
	
	
	
	\bibitem{Shen:siam2010} 
	Shen J, Yang X. A phase-field model and its numerical approximation for two-phase incompressible flows with different densities and viscosities. SIAM  J Sci Comput 2010;32(3):1159-1179.
	
	\bibitem{Cahn:jcp1958} 	
	Cahn JW, Hilliard JE. Free energy of a nonuniform system. I. Interfacial free energy.  J. Chem. Phys. 1958;28(2):258-267.
	
\bibitem{Lowengrub:prs1998}

Lowengrub J, Truskinovsky L. Quasi–incompressible Cahn–Hilliard fluids and topological transitions. Proc. Roy. Soc. London A 1998;454:2617-2654.
	
	
	
	
	
	
\bibitem{Badalassi:jcp2003} 
Badalassi VE, Ceniceros HD, Banerjee S. Computation of multiphase systems with phase field models. J. Comput. Phys. 2003;190(2): 371-397.


\bibitem{Kim:jcp2004} 
Kim J, Kang K, Lowengrub J. Conservative multigrid methods for Cahn–Hilliard fluids. J. Comput. Phys. 2004;193(2):511-543.

\bibitem{Gomez:cmame2008} 
Gómez H, Calo VM, Bazilevs Y, Hughes T. Isogeometric analysis of the Cahn–Hilliard phase-field model. Comput. Methods Appl.Mech. Engrg. 2008;197(49-50):4333-4352.
	
	
	
	\bibitem{Worner:mn2012} 
	Wörner M. Numerical modeling of multiphase flows in microfluidics and micro process engineering: a review of methods and applications.  Microfluid. Nanofluid. 2012;12(6):841-886.
	
	
	
		\bibitem{Savina:pre2003}
	Savina TV, Golovin AA, Davis SH, Nepomnyashchy AA. Faceting of a growing crystal surface by surface diffusion. Phys. Rev. E 2003;67(2):021606.
		
	\bibitem{Wise:jcp2007} 	
	Wise S, Kim J, Lowengrub J. Solving the regularized, strongly anisotropic Cahn–Hilliard equation by an adaptive nonlinear multigrid method. J. Comput. Phys. 2007;226(1):414-446.
	
	\bibitem{Torabi:mpes2009} 
Torabi S, Lowengrub J, Voigt A, Wise, S. A new phase-field model for strongly anisotropic systems.  Proc. R. Soc. A Math. Phys. Eng. Sci. 2009;465(2105):1337-1359.
	

	
	\bibitem{Chen:ccp2013} 	
Chen F, Shen J. Efficient energy stable schemes with spectral discretization in space for anisotropic Cahn-Hilliard systems. Commun.Comput.Phys. 2013;13(5):1189-1208.
	

	
	\bibitem{Chen:cmame2019} 	
Chen C, Yang X. Fast, provably unconditionally energy stable, and second-order accurate algorithms for the anisotropic Cahn–Hilliard model.  Comput. Methods Appl. Mech. Engrg. 2019;351:35-59.

    \bibitem{Fakhari:cnsns2009} 		
Fakhari A, Rahimian MH. Simulation of falling droplet by the lattice Boltzmann method. Commun. Nonlinear Sci. Numer. Simul. 2009; 14(7):3046-3055.
 
	\bibitem{Huang:2015} 	
Huang H, Sukop M, Lu X. Multiphase lattice Boltzmann methods: Theory and application. 2015.
	
	\bibitem{Li:pecs2016} 	
Li Q, Luo KH, Kang QJ, He YL, Chen Q,  Liu Q. Lattice Boltzmann methods for multiphase flow and phase-change heat transfer. Prog. Energy Combust. Sci. 2016;52:62-105.


    \bibitem{Wang:cnsns2024} 
Wang Y, Xiao X, Feng X. Numerical simulation for the conserved Allen–Cahn phase field model of two-phase incompressible flows by an efficient dimension splitting method. Commun. Nonlinear Sci. Numer. Simul. 2024;131:107874.

	\bibitem{He:jcp1999}
He X, Chen S, Zhang R. A lattice Boltzmann scheme for incompressible multiphase flow and its application in simulation of Rayleigh–Taylor instability. J. Comput. Phys. 1999;152(2):642-663.
	
	\bibitem{Mukherjee:pre2007} 	
Mukherjee S, Abraham J. Lattice Boltzmann simulations of two-phase flow with high density ratio in axially symmetric geometry. Phys. Rev. E 2007;75(2):026701.
	
	\bibitem{Lee:jcp2010} 	
Lee T, Liu L. Lattice Boltzmann simulations of micron-scale drop impact on dry surfaces. J. Comput. Phys. 2010;229(20):8045-8063.
	\bibitem{Liang:pre2014} 
Liang H, Shi BC, Guo ZL, Chai ZH. Phase-field-based multiple-relaxation-time lattice Boltzmann model for incompressible multiphase flows. Phys. Rev. E 2014;89(5):053320.
	
	\bibitem{Liang:cnsns2020} 
Liang H, Zhang C, Du R, Wei Y. Lattice Boltzmann method for fractional Cahn-Hilliard equation. Commun. Nonlinear Sci. Numer. Simul. 2020;91:105443.
	
	

\bibitem{Jacqmin:jcp1999} 	
Jacqmin D. Calculation of two-phase Navier-Stokes flows using phase-field modeling. J Comput Phys 1999;155:96–127.
\bibitem{Lee:cpc2012} 
Lee HG, Kim J. An efficient and accurate numerical algorithm for the vector-valued Allen–Cahn equations. Comput. Phys. Commun. 2012;183(10):2107-2115.	
	

	
	
	\bibitem{Wang:C2019} 
	Wang H, Yuan X, Liang H, Chai Z, Shi B.  A brief review of the phase-field-based lattice Boltzmann method for multiphase flows. Capillarity, 2019;2(3):33-52.

	\bibitem{Sekerka:JCG2005} 
Sekerka RF, Analytical criteria for missing orientations on three-dimensional equilibrium shapes, J. Cryst. Growth 2005;(275):77–82
		
\bibitem{Shen:MMAMS2012} 
Shen J. Modeling and numerical approximation of two-phase incompressible flows by a phase-field approach. Multiscale modeling and analysis for materials simulation. 2012;147-195.


\bibitem{He:pre1997} 
 He X, Luo L S. Theory of the lattice Boltzmann method: From the Boltzmann equation to the lattice Boltzmann equation. Phys.
 Rev. E 1997;56(6):6811.

\bibitem{Ginzburg:ccp2008} 
 Ginzburg I, Verhaeghe F, d’Humieres D. Two-relaxation-time lattice Boltzmann scheme: About parametrization, velocity, pressure and mixed boundary conditions. Commun. Comput. Phys. 2008;3(2):427-478.
 

\bibitem{Lallemand:pre2000}
Lallemand P, Luo LS. Theory of the lattice Boltzmann method: Dispersion, dissipation, isotropy, Galilean invariance, and stability. Phys. Rev. E 2000;61(6):6546.

\bibitem{Luo:pre2011}
Luo LS, Liao W, Chen X, Peng Y,  Zhang W. Numerics of the lattice Boltzmann method: Effects of collision models on the lattice Boltzmann simulations. Phys. Rev. E 2011;83(5):056710.

\bibitem{Chai:siam2019} 
Chai Z, Liang H, Du R, Shi B. A lattice Boltzmann model for two-phase flow in porous media. SIAM J. Sci. Comput. 2019;41(4):B746-B772.


\bibitem{Bao:pre2024} 
Bao J, Guo Z. Phase-field lattice Boltzmann model with singular mobility for quasi-incompressible two-phase flows. Phys. Rev. E  2024;109(2):025302.
\bibitem{dHumieres:p2002} 
d'Humières D. Multiple–relaxation–time lattice Boltzmann models in three dimensions. Philos. Trans. R. Soc. Lond. A  2002;360(1792):437-451.

\bibitem{Guo:pre2011}
Guo Z, Zheng C, Shi B. Force imbalance in lattice Boltzmann equation for two-phase flows. Phys. Rev. E 2011;83(3):036707.

	
\bibitem{Lou:el2012}	
Lou Q, Guo Z, Shi B. Effects of force discretization on mass conservation in lattice Boltzmann equation for two-phase flows. Europhys. Let. 2012; 99(6):64005.	

\bibitem{Spencer:pre2004}
Spencer BJ. Asymptotic solutions for the equilibrium crystal shape with small corner energy regularization. Phys. Rev.  2004;69(1):011603.


\bibitem{Maier:pof1996}
Maier RS, Bernard RS, Grunau DW. Boundary conditions for the lattice Boltzmann method. Phys. Fluids 1996;8(7):1788-1801.
	
\end{thebibliography}

\end{document}